# The Agentic Economy: Humans, AI Agents, Robots, and the Measurable Transition toward Distributed Economic Action

**Davit Gondauri,**
**ORCID: https://orcid.org/0000-0002-9611-3688**
PhD, Professor, Business & Technology University, Georgia

**Mikheil Batiashvili,**
**ORCID: https://orcid.org/0009-0004-7752-0423**
Professor, Chairman of Supervisory board, Business & Technology University, Georgia

### Abstract

This article develops the concept of the agentic economy and empirically diagnoses its measurable preconditions: an emerging transition in which economic action is increasingly distributed among humans, AI agents, industrial robots, executable protocols, compute infrastructures and energy systems. The paper argues that classical categories such as labour, capital, firm, market, productivity and trust remain essential, but are no longer sufficient when technologies do not merely raise productivity but also prepare decisions, coordinate workflows, support or execute tasks, verify transactions and reshape responsibility. Methodologically, the study applies a conceptual-empirical quantitative diagnostic design rather than a causal econometric model. It uses public institutional data on AI investment, AI adoption, industrial robot installations and operational stock, data-centre electricity demand and labour-market reallocation. The reported values are transformed through transparent calculations including percentage-point change, relative growth, compound annual growth rate, growth multiplier, stock-flow ratio, concentration ratio and Herfindahl-Hirschman Index. The results indicate that the measurable preconditions of the agentic economy are already visible: AI adoption is accelerating in official and organisational indicators; AI investment is a broad capital-allocation signal rather than a direct measure of new productive capital allocation; industrial robots are treated not as full autonomous agents but as persistent cyber-physical action capacity; AI-related compute expansion contributes to wider data-centre electricity pressure; and available labour-market projections are more consistent with task reallocation than with simple labour disappearance. The article contributes a new action-capacity framework that links model/software-agent capacity, robotic capacity, compute-energy coupling, protocolisation, auditable trust and human sovereignty. The findings support a disciplined claim: the agentic economy is not yet a completed global institutional order, and the present article does not directly measure full agentic delegation; rather, its transition pressure is sufficiently measurable to require a distinct economic vocabulary, a reproducible diagnostic methodology and future sector-level measurement of coordination friction, auditability, protocol failure and human override capacity.

**Keywords:** agentic economy; AI agents; robotics; action capacity; coordination friction; compute-energy coupling; auditable trust; human sovereignty; labour reallocation; digital transformation

**JEL Classification**: O33, L86, J24, D83, D24, Q43, L23, O14.

## 1. Introduction

### 1.1. Background and research problem

The contemporary economy is undergoing a transformation that cannot be fully captured by the conventional vocabulary of technology adoption, digitalisation or automation. The central change is deeper: the bearer of economic action is changing. In industrial capitalism, the dominant unit of action was the human worker organised inside the firm, using physical capital and exchanging through markets. In the digital economy, platforms, data infrastructures and algorithmic systems expanded the scope of coordination. In the emerging agentic economy, however, economic action is increasingly performed, prepared, mediated, verified and governed by a mixed architecture of humans, AI agents, robots, protocols, compute infrastructure and energy-dependent systems. This article begins from that structural observation and asks whether the measurable preconditions of such an economy are already visible in contemporary global data.

The concept of an agentic economy is not used here as a rhetorical replacement for the digital economy. It is proposed as a more precise analytical category for a specific stage of technological and institutional change. A digital system stores, transmits and processes information; a platform intermediates and governs networked interactions; an automated system substitutes or augments particular tasks. An agentic system goes further when it participates in the selection, sequencing, execution, monitoring, verification or correction of economic action under rule-based and auditable conditions. This distinction is essential because an economy may be highly digital without being fully agentic, and a firm may use AI tools without delegating economically meaningful agency to software agents or protocols.

The problem addressed by the article is therefore not whether artificial intelligence and robotics exist or whether firms use them. That is already empirically evident. The more important research problem is whether these technologies create a need to expand the fundamental economic categories through which production, coordination, capital, labour, trust and governance are analysed. If models can support or prepare decisions, if software agents can execute tasks, if robots can act in physical space, if protocols can determine access and settlement, and if compute systems depend materially on energy infrastructure, then the economy is no longer adequately described by treating technology only as a neutral productivity shifter. The article proposes that economic theory must also analyse action capacity: the capacity of a socio-technical system to act reliably, efficiently, verifiably and legitimately.

The article is written as a conceptual-empirical quantitative diagnostic study. Its goal is not to estimate a causal effect of AI or robotics on GDP, employment or productivity. Such a causal design would require harmonised firm-level or sector-level datasets on delegated AI tasks, robot usage, protocolised settlement, audit-log completeness, model-error costs, latency and human override. Such data are not yet available globally. Instead, the paper tests whether observable indicators in AI investment, AI adoption, industrial robot stock, data-centre electricity demand and labour-market projections are sufficiently strong and convergent to justify a new analytical framework. This makes the contribution both empirical and conceptual: empirical because it uses real institutional statistics and transparent calculations, and conceptual because it connects those results to the theory of distributed economic action.

### 1.2. Why classical economic language remains necessary but incomplete

The argument does not reject classical or institutional economics. On the contrary, it begins from their strongest insights. Coase (1937) showed that the firm exists partly because using the price mechanism is costly. Hayek (1945) explained how markets coordinate dispersed knowledge through prices. Simon (1955) demonstrated the importance of bounded rationality. Williamson (1975) and North (1990) showed that governance structures and institutions shape economic performance. These foundations remain indispensable. However, the institutional question must now be extended: if search, matching, contracting, monitoring, compliance and settlement can be partly executed by AI systems and protocols, the nature of coordination costs changes.

The same applies to production theory. The classical production function, in which output depends on capital, labour and technology, remains useful for many macroeconomic and productivity questions. Yet it compresses too much into the technology term when models, robots, compute, protocols and auditability become separate determinants of action. Solow (1957) placed technical change at the centre of growth accounting, while Romer (1990) and Bresnahan and Trajtenberg (1995) developed the importance of knowledge, innovation and general-purpose technologies. The agentic economy builds on this tradition but adds that some technologies are no longer only efficiency parameters; they become components of action architecture.

The old vocabulary also struggles with trust. In traditional economic analysis, trust is associated with reputation, contract, law and institutional enforcement. In the agentic economy, trust increasingly requires logs, traceability, model documentation, contestability, explainability, audit trails and override procedures. A transaction executed through software or protocol infrastructure may be fast and efficient but still economically dangerous if it is not auditable. This is why the article treats auditable trust as a distinct component of action capacity rather than as a secondary governance issue.

The incompleteness of the old vocabulary is therefore not a failure of economics. It is a historical limit of categories that were designed for a world in which action was primarily human, machines were primarily tools, and markets were primarily price systems. The new environment requires an extension rather than abandonment. Labour must be reread as execution, coordination, information, audit and sovereign judgement. Capital must include physical capital, model capital, compute infrastructure and robotic capital. Markets must be analysed as price mechanisms and protocol-mediated coordination systems. Productivity must be supplemented by coordination-adjusted and energy-adjusted performance.

### 1.3. From digitalisation and automation to the agentic economy

Digitalisation, automation, platformisation and agentic transformation overlap, but they should not be conflated. Digitalisation refers to the conversion of processes and information into digital forms. Automation refers to the substitution or augmentation of tasks by machines or software. Platformisation refers to the organisation of markets and interactions through digital intermediaries, network effects and ecosystem governance. Agentic transformation begins when economic action itself is redistributed across humans, AI agents, robots and executable protocols. This is why the article does not define the agentic economy only by the presence of digital tools, but by the redistribution of action rights, execution capacity, verification responsibility and human override.

Platform economics helps explain the institutional background of this shift. Rochet and Tirole (2003) showed that multi-sided platforms structure participation through cross-side network effects. Gawer (2014) and Parker, Van Alstyne and Choudary (2016) further developed the platform as an architecture of economic coordination. The agentic economy extends this analysis by focusing on executable coordination. A platform can rank, match and intermediate; a protocol can predefine access, automate settlement, enforce compliance, record logs and shape the conditions under which transactions occur. Therefore, the market is not abolished by agentic systems; it is increasingly embedded inside programmable mechanisms.

Artificial intelligence intensifies this shift because it can become part of organisational cognition. Agrawal, Gans and Goldfarb (2018) interpret AI as a technology that reduces the cost of prediction, thereby increasing the value of judgement and complementary decision structures. Foundation-model literature shows that large models can support language, coding, classification and reasoning-like tasks (Bommasani et al., 2021; Brown et al., 2020; OpenAI, 2023). When such systems are embedded in workflows, they become more than technical tools. They become components of organisational action capacity.

Robotics gives the transformation a physical dimension. Industrial robots are not treated here as full autonomous agents; they are interpreted more cautiously as persistent cyber-physical action capacity. They remain economically important because they move, assemble, transport, inspect, stop, fail, require maintenance, consume energy and create safety and liability considerations. The agentic economy is therefore not purely digital. It is simultaneously informational, physical, infrastructural and institutional. This is why the article uses both AI indicators and robot-stock indicators: software-agent capacity and robotic embodiment are complementary forms of non-human economic action, while the degree of autonomy of each system must be assessed empirically rather than assumed.

### 1.4. Empirical motivation and measurable transition pressure

The empirical motivation of the article comes from the simultaneous movement of several data families. First, AI investment has reached a scale that makes it a broad capital-allocation signal. The manuscript uses Stanford HAI AI Index data reporting corporate AI investment of USD 252.3 billion in 2024 and generative-AI private investment of USD 33.9 billion. These figures are interpreted not merely as technology expenditure but as investment in model capability, compute access, data infrastructure, talent and organisational standards.

Second, AI adoption is now visible in both official and survey-based indicators. Eurostat enterprise statistics, OECD firm-level data and McKinsey organisational survey evidence point to expanding AI use. The study treats these sources with comparability discipline: they are not merged into a single panel because they measure different populations and definitions. Instead, they are used as convergence evidence that AI is entering the internal structure of organisations. This is the empirical reason why the article distinguishes AI use, AI integration and agentic transformation.

Third, industrial robotics demonstrates the cyber-physical side of the transition. The Results chapter uses International Federation of Robotics data on industrial robot installations and operational stock. These data allow the article to calculate a stock-flow ratio and interpret robots as a persistent stock of embodied action capacity rather than a temporary annual investment flow. Regional concentration in robot installations is also meaningful because it shows that future industrial competitiveness may depend not only on labour cost, but also on robotic throughput, maintenance capacity, safety standards and industrial energy reliability.

Fourth, compute-energy coupling introduces a material boundary to AI economics. The International Energy Agency's data-centre electricity estimates and projections are used to show that AI and compute cannot be analysed independently of electricity, grid connection, cooling, location and energy security. A country or firm may adopt AI tools, but if its energy infrastructure is weak, its real agentic capacity will be constrained. This makes energy an internal variable of the agentic economy rather than an external background condition.

Fifth, labour-market projections indicate that the transition should not be read through a simple disappearance narrative. The World Economic Forum projection used in the article reports large gross flows of new and displaced roles by 2030. The article interprets this through the task-based labour literature (Acemoglu & Autor, 2011; Autor et al., 2003; Frey & Osborne, 2017): the central question is not whether labour vanishes, but how tasks are repartitioned across execution, coordination, information, audit and sovereign functions.

### 1.5. Research gap and theoretical contribution

Existing literature provides strong but fragmented foundations. Institutional economics explains coordination and governance, but not the full distribution of action across AI agents, robots and protocols. Innovation economics explains general-purpose technologies and complementary investments, but often treats technology as a productivity driver rather than as an acting infrastructure. Platform economics explains networked intermediation and gatekeeping, but does not always distinguish platform power from executable protocol power. Labour economics explains task reallocation under automation, but less often incorporates audit labour and sovereign labour. AI governance literature explains accountability and oversight, but usually outside a unified economic model of action capacity. Energy research explains data-centre electricity demand, but is often separated from the core language of economic coordination.

The article addresses this gap by integrating these literatures around the concept of action capacity. Action capacity refers to the ability of a socio-technical economic system to select, execute, verify, correct and govern action. This category is broader than productivity because it includes speed, reliability, auditability, legitimacy and infrastructure feasibility. It is also broader than automation because it includes human sovereignty and governance. The proposed action-capacity function identifies human judgement, conventional capital, model/software-agent capacity, robotic capacity, protocol quality, compute capacity, energy availability, auditable trust and institutional uncertainty as analytically distinct variables.

A second theoretical contribution is coordination friction. The article extends transaction-cost economics by incorporating new forms of friction associated with agentic systems: latency, model error, protocol failure, cyber-physical risk, audit gaps, energy bottlenecks and human-override burden. This concept is important because agentic systems can reduce old costs while creating new ones. For example, AI may reduce search and documentation time but introduce model-error and auditability costs; robotics may raise throughput but increase maintenance and energy dependence; protocols may reduce settlement delay but create new failure points and exclusion risks.

A third contribution is the concept of human sovereignty. Human sovereignty does not mean that humans perform every task. It means that humans retain responsibility for goals, normative boundaries, exception interpretation and final override. This concept is consistent with AI governance frameworks such as the OECD AI Principles, NIST AI Risk Management Framework and the EU Artificial Intelligence Act, but the article places it within economic theory rather than treating it only as an ethical add-on. A system that acts rapidly but cannot be audited, contested or overridden may be efficient in a narrow sense while being economically and institutionally fragile.

### 1.6. Research aim, research questions and hypotheses

The main aim of the article is to develop the concept of the agentic economy and empirically diagnose its measurable preconditions as an emerging transition toward distributed economic action across humans, AI agents, robots, protocols and compute-energy infrastructures. The article does not aim to prove that the global economy has already completed this transformation. Its aim is more precise: to test whether measurable transition pressure is already visible in institutional statistics and whether this pressure justifies a new analytical vocabulary.

The study is guided by five research questions. First, are the measurable preconditions of the agentic economy visible across AI adoption, investment, robotics, energy and labour-market indicators? Second, which dimensions show the clearest transition pressure: AI adoption, AI capital allocation, compute-energy demand, robotic embodiment or labour reallocation? Third, is the agentic economy only a digital phenomenon, or does robotics show a cyber-physical extension of economic action? Fourth, does the labour evidence support a disappearance narrative or a reallocation/repartition narrative? Fifth, do the empirical indicators justify the claim that the old economic vocabulary is under-specified for distributed action?

The hypotheses are diagnostic rather than causal. H1 proposes that firm-level AI adoption is accelerating. H2 proposes that AI-related compute expansion contributes to wider data-centre electricity pressure. H3 proposes that robotic embodiment is already a global industrial reality, but industrial robots are treated as cyber-physical action capacity rather than full autonomous agents. H4 proposes that available labour-market projections are more consistent with a reallocation narrative than with a simple disappearance narrative. P5 is treated not as an econometric hypothesis, but as a theoretical proposition: the convergence of measured indicators supports the proposition that classical economic categories are under-specified for distributed socio-technical action. H1-H4 are evaluated using transparent calculations and evidence boundaries rather than p-values; P5 is assessed as a conceptual proposition supported by empirical convergence. This is appropriate because the study is not designed as a causal econometric test.

The research design therefore fits the maturity level of the field. A premature regression model could create false precision by merging incomparable sources or treating projections as realised outcomes. A diagnostic model, by contrast, makes visible the structure and strength of transition pressure while preserving methodological caution. This is the central reason why the article is framed as a quantitative diagnostic study.

### 1.7. Methodological logic and empirical discipline

The methodology uses externally reported values from recognised institutional sources and transforms them only through transparent calculations: absolute change, percentage-point change, relative change, compound annual growth rate, growth multiplier, scale ratio, stock-flow ratio and Herfindahl-Hirschman Index. No hypothetical numerical values are introduced in the Results section. This makes the empirical design reproducible and reduces the risk of speculative quantification, a frequent weakness in emerging-technology studies.

The methodology also separates evidence classes. Observed or reported facts include AI investment, enterprise AI use, robot installations, operational robot stock and 2024 data-centre electricity use. Institutional projections include IEA data-centre electricity projections and WEF labour-market projections. Author-calculated indicators include CAGR, growth multiplier, stock-flow ratio and HHI. Theoretical interpretation includes the mapping of measured indicators to action capacity, coordination friction, robotic embodiment, compute-energy coupling and human sovereignty. This classification is essential because it prevents the article from presenting projections as realised facts or conceptual interpretations as measured variables.

The empirical design is intentionally multi-domain. The agentic economy cannot be measured by one indicator because its components are distributed across investment, adoption, industrial automation, energy and labour. A single aggregate index could hide this heterogeneity. The article therefore first reports the separate indicators and then maps them to the theoretical framework. This approach makes the logic of the argument visible: AI adoption and investment correspond to model/software-agent capacity; robot stock corresponds to robotic capacity; data-centre electricity demand corresponds to compute-energy coupling; labour-market reallocation corresponds to human role transformation; and protocol quality and auditable trust are identified as future sector-level measurement priorities.

This methodological discipline is intended to make the article suitable for peer review. It avoids unsupported causal claims, avoids merging non-comparable adoption series, marks projections as projections, and states the limitations of global aggregate data. At the same time, it does not retreat into purely descriptive commentary. It uses real statistics to support a theoretical claim about the changing architecture of economic action.

### 1.8. Scope, boundaries and limitations

The scope of the article is global and diagnostic. It does not provide a country-specific index of agentic readiness, nor does it estimate firm-level performance effects. It also does not measure all components of the proposed action-capacity framework directly. Protocol quality, auditability and human override capacity are theoretically central but require sector-level operational data that are not yet available in global public datasets. This is why the article identifies them as future measurement priorities rather than treating them as fully observed variables.

Another boundary concerns adoption. The article does not equate AI adoption with full agentic transformation. AI use can be limited to one business function; AI integration can be recurring but still human-directed; full agentic transformation requires delegated preparation, execution, verification or coordination under auditability and override conditions. This distinction is necessary because otherwise the article would overstate the meaning of adoption statistics.

A further limitation concerns projections. IEA and WEF projections are useful because they indicate expected pressure on energy systems and labour markets, but they are not realised outcomes. The article therefore interprets them as institutional scenarios. This conservative interpretation is methodologically important for maintaining the distinction between observed facts and future expectations.

These limitations do not weaken the article if they are stated clearly. They define the empirical boundary of the study. The article claims that measurable transition pressure is visible, not that the agentic economy is complete in every firm, sector or country. This makes the argument strong but defensible: the new vocabulary is justified by empirical convergence, while the full measurement of agentic transformation remains a future research programme.

### 1.9. Structure of the article

The article proceeds in six chapters. Chapter 1 introduces the research problem, the concept of the agentic economy, the empirical motivation, the research questions and the methodological boundary. Chapter 2 reviews the relevant literature across institutional economics, innovation economics, platform economics, labour economics, AI research, robotics, energy economics and AI governance. Chapter 3 develops the methodology, including the diagnostic research design, action-capacity function, coordination-friction architecture, variable operationalisation, formulas, algorithms and reproducibility logic.

Chapter 4 presents the Results and Empirical Diagnostics. It reports the empirical matrix, comparability controls, AI adoption acceleration, AI capital concentration, compute-energy coupling, robotic embodiment, labour-market reallocation, mapping to the action-capacity framework, hypothesis assessment and calculation audit. Chapter 5 interprets the findings as transition pressure rather than deterministic proof and explains their theoretical, methodological and policy implications. Chapter 6 concludes the article by synthesising the theoretical contribution, empirical findings, limitations and future research agenda.

## 2. Literature Review

The literature relevant to the agentic economy is distributed across several traditions rather than contained in one established field. Classical and institutional economics explain coordination, transaction costs, dispersed knowledge, bounded rationality and governance; innovation economics explains general-purpose technologies and intangible complementary capital; platform economics explains digitally mediated market architecture; labour economics explains task reallocation under automation; robotics economics connects machine action to industrial capital; energy economics explains the material boundary of compute; and AI-governance scholarship explains auditability, accountability, safety and human oversight. This review synthesises these streams into a single argument: the agentic economy is not simply a new label for digitalisation, automation or platformisation, but an emerging coordination order in which economic action is redistributed across humans, AI agents, robots, protocols, data infrastructures and energy-dependent compute systems.

The chapter is designed to support the article's diagnostic methodology and empirical results. The Results section measures AI adoption, AI investment concentration, robotic operational stock, data-centre electricity pressure and labour-market reallocation; the Interpretation chapter reads those findings as transition pressure rather than deterministic proof. The literature review therefore has a specific function: it must show why these indicators belong together theoretically. It builds the bridge from old categories—labour, capital, firm, market, technology and trust—to the article's proposed categories of action capacity, coordination friction, compute-energy coupling, robotic embodiment, auditable trust and human sovereignty.

### 2.1. Institutional foundations: coordination, transaction costs and the limits of the old vocabulary

The first foundation of the agentic-economy framework is institutional economics. Coase (1937) showed that the firm exists partly because using the price mechanism is costly; economic coordination is therefore never frictionless. Hayek (1945) emphasised that markets coordinate dispersed knowledge, while Simon (1955) showed that decision-makers operate under bounded rationality rather than perfect optimisation. Arrow (1974), Williamson (1975) and North (1990) then deepened this institutional view by linking organisation, governance, rules and institutional constraints to economic performance. These works remain essential because the agentic economy is also a coordination problem. The difference is that the instruments of coordination have changed: price, contract and hierarchy are now joined by software agents, APIs, automated compliance systems, robotic fleets, platform rules and auditable logs.

This literature makes clear why a purely production-centred or price-centred economic language is insufficient. If transaction costs shape the boundary between firm and market, then the automation of search, contracting, monitoring, settlement and verification changes the firm-market boundary itself. In the agentic economy, coordination can be executed by programmable rules and model-driven systems rather than only by human managers or market prices. The literature on institutions therefore supports the article's concept of coordination friction: the relevant cost field includes not only search, bargaining and enforcement, but also latency, model error, protocol failure, audit gaps, cyber-physical risk, energy interruption and human-override burden.

The same institutional tradition also explains why the article does not treat AI adoption as automatically beneficial. Institutions govern how technology is used, who controls it, who can contest its outputs and who captures the resulting surplus. North's (1990) emphasis on institutional constraints is especially relevant: the same AI or robotics technology can produce inclusive augmentation in one institutional setting and concentrated protocol rent in another. This is why the article links empirical indicators to human sovereignty and auditability rather than to technological diffusion alone.

### 2.2. Innovation, productivity and general-purpose technologies

A second stream concerns technological change and productivity. Solow (1957) formalised the importance of technical change for growth, while Romer (1990) made knowledge and endogenous innovation central to long-run economic development. Nelson and Winter (1982) introduced an evolutionary view in which firms learn through routines, capabilities and selection processes. Bresnahan and Trajtenberg (1995) framed general-purpose technologies as engines of growth whose impact depends on complementary innovation across sectors. This literature is directly relevant because AI is often interpreted as a general-purpose technology, but the agentic-economy framework adds a further step: AI is not only a productivity shifter; it can participate in the preparation, execution, monitoring and verification of economic action.

The productivity literature also warns against expecting immediate macroeconomic gains from technological breakthroughs. Brynjolfsson and Hitt (2000) showed that information technology produces performance effects through organisational transformation rather than hardware alone. Jorgenson (2001) documented the macroeconomic importance of information technology investment, but later work on the productivity paradox emphasised delayed measurement effects and complementary intangible capital. Brynjolfsson, Rock and Syverson (2019, 2021) argued that new general-purpose technologies may first appear as weak productivity statistics because firms must redesign processes, build intangible assets and reorganise work before gains become visible. The agentic economy is consistent with this argument: AI adoption becomes economically meaningful only when it is embedded in workflows, governance, training, audit and infrastructure. The article's empirical design follows this tradition by refusing to infer productivity effects directly from adoption figures. Its Results section measures transition pressure, not final productivity. This is consistent with Varian's (2010) view that computer-mediated transactions change economic measurement and with the broader information-economy literature that treats digital systems as reconfigurations of transaction architecture. Therefore, AI investment, firm-level AI use and data-centre electricity demand are not interpreted as automatic proof of growth; they are interpreted as observable preconditions for a deeper organisational transformation.

### 2.3. Digital, platform and protocol-mediated coordination

The digital and platform-economy literature supplies the third foundation. Rochet and Tirole (2003) formalised two-sided platform markets, showing that pricing and participation depend on cross-side network effects. Gawer (2014) developed an integrative view of technological platforms, while Tiwana (2014) connected platform architecture to governance and strategy. Parker, Van Alstyne and Choudary (2016) popularised the platform model as an economic architecture built around networked participation. Kenney and Zysman (2016) described the rise of the platform economy, and Van Dijck, Poell and de Waal (2018) analysed platforms as institutional infrastructures that organise public values as well as markets. This literature is crucial but not sufficient for the agentic economy. Platform economics explains network effects, gatekeeping and ecosystem control. The agentic framework asks an additional question: when does a platform or protocol cease to be only an intermediary and become an executable architecture of action? A platform may rank, match, price and monitor; a protocol may go further by predefining access, compliance, settlement, identity verification and auditability. The agentic economy therefore extends platform economics toward protocol-mediated coordination: markets are no longer only price systems but rule-execution environments.

This distinction directly supports the article's empirical interpretation of AI capital concentration. Investment in models, data pipelines, cloud infrastructure and deployment ecosystems is not merely investment in a productive input. It may indicate emerging coordination-power asymmetries, especially where model access, compute infrastructure, cloud ecosystems and deployment standards are concentrated. The platform literature helps explain how network effects generate concentration; the agentic-economy framework adds that executable rules and model-dependent workflows can generate protocol rent and dependence. For this reason, the article interprets AI investment concentration as a potential source of coordination-power asymmetry rather than as ordinary capital accumulation alone.

### 2.4. Automation, tasks and labour-market reallocation

The labour literature provides the fourth pillar. Autor, Levy and Murnane (2003) shifted automation analysis from occupations to tasks, showing that computerisation affects routine and non-routine work differently. Acemoglu and Autor (2011) further developed the task-based framework, while Autor (2015) argued that automation has historically displaced some tasks while complementing others and creating new demand. This task-based perspective is essential for the article because the agentic economy is not primarily about the disappearance of occupations; it is about the redistribution of tasks across humans, AI agents, robots and protocols.

The automation-risk literature deepens this point. Frey and Osborne (2017) estimated susceptibility of jobs to computerisation, while Arntz, Gregory and Zierahn (2016) showed that task-based estimates often imply lower automation risk than occupation-level estimates. Acemoglu and Restrepo (2018, 2019, 2020) developed a broader framework in which automation displaces tasks, but new tasks may reinstate labour demand; their empirical work on robots documents labour-market effects while also showing the importance of technology-labour interaction. Graetz and Michaels (2018) demonstrated that industrial robots can raise productivity and value added, but the distributional effects depend on sectoral and labour-market structures.

Recent AI-specific work makes the task perspective even more important. Felten, Raj and Seamans (2021) developed measures of occupational exposure to AI, while Webb (2020) constructed patent-text-based exposure measures for AI, software and robots. Bessen (2018) argued that demand effects matter when analysing the employment consequences of AI. Agrawal, Gans and Goldfarb (2018) interpreted AI as a fall in the cost of prediction, shifting the value of judgement, data and complementary decisions. Brynjolfsson, Li and Raymond (2023), Noy and Zhang (2023), and Dell'Acqua et al. (2023) show that generative AI can significantly affect knowledge-worker productivity, but also that outcomes depend on task structure, expertise, organisational design and the boundaries of the technology. Susskind (2020) highlights the deeper policy question: how institutions should respond if labour-market adjustment becomes structurally uneven.

This literature justifies the article's labour interpretation. The WEF projection of simultaneous job creation and displacement is not read as a simple net-employment statistic. It is read through the task literature as evidence of labour repartition. The relevant categories become execution labour, coordination labour, information labour, audit labour and sovereign labour. Human labour does not disappear from the economic system; it shifts toward goal-setting, supervision, exception interpretation, contestability and legitimate override. OECD labour-market work similarly supports the need for careful institutional interpretation rather than deterministic automation narratives.

### 2.5. Artificial intelligence as agent capacity: models, prediction and foundation systems

The AI literature provides the technical basis for understanding why software systems can become economically agentic. Russell and Norvig (2021) define AI around intelligent agents that perceive and act, while Russell (2019) frames the problem of control as central to safe AI deployment. Goodfellow, Bengio and Courville (2016) and LeCun, Bengio and Hinton (2015) establish the deep-learning foundation for contemporary AI capabilities. Scaling-law research by Kaplan et al. (2020) and Hoffmann et al. (2022) shows that model performance is linked to data, parameters and compute; this makes compute capacity an economic infrastructure rather than a purely technical detail.

Foundation-model research strengthens the article's argument that AI is becoming a general organisational layer. Brown et al. (2020) demonstrated few-shot learning in large language models; Bommasani et al. (2021) described foundation models as a paradigm with broad opportunities and risks; OpenAI (2023) documented the capabilities and limitations of GPT-4. These sources support the article's distinction between ordinary software and model/software-agent capacity. A foundation model is not simply a fixed tool: when embedded into workflows, it can generate text, classify information, support decisions, plan actions and interact with other systems. It therefore becomes part of action capacity.

However, the same AI literature supports methodological caution. The existence of powerful foundation models does not prove that firms have become agentic. AI use, AI integration and agentic transformation are different levels. AI use may involve isolated tools; integration involves recurring workflows; agentic transformation requires delegated preparation, execution, verification or coordination under auditability and human override. This is why the article's Results section treats McKinsey, Eurostat and OECD adoption indicators as signals of transition pressure, not as proof of complete agentic transformation.

### 2.6. Robotics, cyber-physical embodiment and industrial agency

Robotics expands the literature review beyond the digital economy. Industrial robots are not merely software systems; they are cyber-physical capital that acts in material space. The International Federation of Robotics (2025) provides the empirical basis for measuring annual installations, operational stock and regional concentration. This source is especially important because it allows the article to distinguish informational AI capacity from embodied robotic capacity. AI systems may support decisions or generate content; robots move, manipulate, inspect, assemble, transport and interact with physical environments.

The robotics literature connects directly to economic studies of automation. Graetz and Michaels (2018) and Acemoglu and Restrepo (2020) show that robots affect productivity, value added and labour markets, but their effects depend on exposure, sectoral structure and institutional adaptation. The agentic-economy framework builds on this by treating robot stock as embodied action capacity. A robot is not only a capital good; it is a physical unit of delegated action that consumes energy, requires maintenance, creates safety and liability risks, and changes the boundary between human execution and machine execution.

This perspective is central to the article's Results and Interpretation chapters. The stock-flow ratio for industrial robots is interpreted as evidence that robotic action has become a persistent global stock, not merely a yearly investment flow. Regional concentration in robot installations is interpreted as uneven distribution of cyber-physical competitiveness. These interpretations are consistent with the literature but extend it by placing robots inside a broader action-capacity framework that also includes AI models, protocols, compute, energy and human sovereignty.

### 2.7. Compute-energy coupling and the material infrastructure of AI

The agentic economy cannot be understood without energy and compute. Data-centre research shows that digital systems are not immaterial. Koomey (2011) and Shehabi et al. (2016) documented the growth and efficiency dynamics of data-centre electricity use. Masanet et al. (2020) recalibrated global data-centre energy estimates and showed the importance of efficiency improvements, while Strubell, Ganesh and McCallum (2019), Schwartz et al. (2020) and Patterson et al. (2021) highlighted the energy and carbon implications of machine-learning models. The International Energy Agency (2025) extends this literature with current projections of AI-related data-centre electricity demand.

This literature directly supports the article's compute-energy coupling concept. If model training and inference require compute, and compute requires electricity, cooling, chips, grid connections and physical infrastructure, then energy must be treated as an internal variable of AI economics. The article therefore separates compute capacity and energy availability in the action-capacity framework. This is methodologically important because two economies may have similar AI adoption rates but very different real agentic capacity if one has reliable energy and data-centre infrastructure while the other faces grid bottlenecks, high electricity costs, connection delays or cooling constraints.

The compute-energy literature also explains why the agentic economy is not only a question of software adoption. Data centres, cloud platforms, chip supply chains, electricity grids and cooling systems become the material foundations of model/software-agent capacity. Robotics reinforces the same conclusion because cyber-physical systems require charging, maintenance, sensor networks and industrial power reliability. This is why the article interprets IEA data-centre projections as infrastructure pressure rather than as a simple digital-growth statistic.

### 2.8. AI governance, auditability and human sovereignty

The fifth major literature stream concerns AI governance and accountability. Floridi and Cowls (2019) propose principles for AI in society, and Jobin, Ienca and Vayena (2019) map the global landscape of AI ethics guidelines. Raji et al. (2020) develop algorithmic auditing as an end-to-end accountability framework. Mitchell et al. (2019) propose model cards for model reporting, and Gebru et al. (2021) propose datasheets for datasets. These contributions are highly relevant because the agentic economy requires not only automation but verifiable action. When systems prepare, recommend, execute or verify economic tasks, accountability depends on documentation, traceability, audit logs, contestability and institutional responsibility.

Policy frameworks point in the same direction. The OECD AI Principles emphasise trustworthy, human-centred AI (OECD, 2019), the NIST AI Risk Management Framework provides a structured approach to AI risk governance (National

Institute of Standards and Technology, 2023), and the EU Artificial Intelligence Act creates a risk-based legal architecture for AI systems (European Parliament & Council of the European Union, 2024). These frameworks support the article's concept of human sovereignty. Human oversight is not enough if the human is only a nominal approver. Sovereignty requires real capacity to set goals, understand risks, interpret exceptions, contest outputs and stop or correct automated systems.

AI-safety literature reinforces this point. Amodei et al. (2016) identify concrete safety problems that arise when AI systems pursue objectives under imperfect specification, while Russell (2019) argues that control and alignment are central to the future of AI. In economic terms, these concerns translate into auditability, override capacity and legitimacy. The agentic economy can increase action capacity, but if it reduces contestability or makes decisions unauditable, it may create systemic risk rather than sustainable value. Therefore, governance is not an external ethical supplement; it is part of the economic production of legitimate action.

### 2.9. Empirical indicator literature: AI adoption, investment, robotics, energy and labour projections

The article's empirical sources are also part of the literature because they define what can currently be measured. Stanford HAI (2025) provides a broad evidence base on AI investment, technical development and economic diffusion. McKinsey and Company / QuantumBlack (2025) provides organisational survey evidence on AI use and the early scaling of agentic AI systems. Eurostat (2025a, 2025b) provides official enterprise-level AI adoption statistics for the European Union, while OECD (2026) provides firm-level adoption evidence across countries with available data. These sources justify the article's use of AI adoption and investment as measurable preconditions of agentic transformation.

Robotics and energy sources provide the second empirical layer. The International Federation of Robotics (2025) provides installation and operational-stock data for industrial robots. The International Energy Agency (2025) provides data-centre electricity estimates and projections, linking AI diffusion to electricity demand and grid planning. The World Economic Forum (2025) provides labour-market projections for job creation, displacement and structural reallocation. Together, these sources support the article's multi-domain diagnostic methodology: no single indicator can capture the agentic economy, but multiple institutional indicators reveal measurable transition pressure.

The use of these institutional sources also creates a methodological boundary. Survey evidence, official enterprise statistics, robot installations, energy projections and labour projections do not have the same evidentiary status. The article therefore uses them as convergent evidence rather than as a single merged dataset. This approach is consistent with the literature's caution about productivity paradoxes, task-level heterogeneity and institutional mediation. The empirical literature supports a strong but disciplined claim: the preconditions of the agentic economy are visible, but full agentic transformation requires future firm-level and sector-level data on delegated workflows, protocolisation, auditability, model error, latency and human override.

### 2.10. The gap in existing literature and the contribution of the agentic-economy framework

The reviewed literature is rich, but it remains fragmented. Institutional economics explains coordination, but not yet AI agents and robots as distributed acting units. Platform economics explains networked intermediation, but not always executable protocols and auditable action. Labour economics explains automation and tasks, but less often the rise of audit labour and sovereign labour. AI governance explains accountability and oversight, but rarely embeds these concepts in a macroeconomic action-capacity model. Energy and data-centre research explains the material cost of AI, but is seldom integrated into the core language of economic coordination. Robotics economics explains embodied automation, but often remains separated from AI-agent and protocol literature.

The article's contribution is to integrate these literatures around the concept of action capacity. The central question becomes not only how resources are allocated, but how economic action is selected, executed, verified, corrected and governed. This changes the meaning of capital, labour, productivity, trust and market coordination. Capital includes model capital, compute infrastructure and robotic capital. Labour includes execution, coordination, information, audit and sovereign labour. Productivity includes coordination-adjusted and energy-adjusted performance. Trust becomes auditable and verifiable trust. Markets become partly protocol-mediated coordination systems.

This synthesis also explains the article's methodological design. Because the agentic economy is emerging, it cannot yet be fully estimated with one causal econometric model. Instead, it can be diagnosed through observable transition-pressure indicators: AI adoption acceleration, AI investment concentration, robotic stock-flow maturity, compute-energy demand growth and labour-market reallocation. This diagnostic approach is not a weakness; it is the appropriate empirical bridge between established literature and a new research object. The next stage of the literature should move toward sector-level measurement of coordination friction, auditability, protocol failure, model-error cost, latency, energy bottlenecks and human-override capacity.

Adjacent literatures further strengthen this gap statement. Algorithmic management literature directly addresses how algorithmic systems allocate, monitor and evaluate work, showing that automated coordination is already reshaping labour control and organisational authority (Kellogg et al., 2020; Lee et al., 2015). Autonomous-agent and multi-agent-systems research provides the technical language for distributed action, interaction, incentives and coordination among artificial actors (Shoham & Leyton-Brown, 2008; Wooldridge & Jennings, 1995). Agent-based computational economics also shows how economic processes can be represented as dynamic systems of interacting agents rather than only as aggregate equations (Tesfatsion, 2006).

Cyber-physical-systems, Industry 4.0 and digital-twin research extend the argument into material production. Cyber-physical systems integrate computation with physical processes (Lee, 2008), Industry 4.0 connects digitalisation to manufacturing architectures (Lasi et al., 2014), and digital twins create model-based links between physical systems and virtual representations for smart manufacturing (Lu et al., 2020). These literatures support the article's cautious interpretation of industrial robots: they are not automatically full autonomous agents, but they do constitute persistent cyber-physical action capacity.

Finally, smart-contract, platform-infrastructure and AI-alignment literatures sharpen the article's governance contribution. Smart contracts show how executable rules can expand the contracting space and transform institutional coordination (Cong & He, 2019). Platform-infrastructure studies explain how digital platforms become infrastructural power, not merely market intermediaries (Plantin et al., 2018). AI-alignment literature reinforces the importance of human values, control and governance in systems capable of consequential action (Gabriel, 2020). These additions make the agentic-economy framework more clearly distinct from ordinary digitalisation, platformisation and automation.

### 2.11. Literature-based propositions for the article

The literature supports five propositions that structure the empirical chapters. First, technology must be analysed institutionally: AI and robots matter not only because they increase efficiency, but because they change coordination, governance and the boundary between firm and market. This proposition follows from Coase, Hayek, Simon, Williamson and North, and is extended by the platform-economy literature.

Second, AI should be treated as a general-purpose and organisational technology whose productivity effects depend on complementary intangible investments. This proposition follows from Romer, Bresnahan and Trajtenberg, Brynjolfsson and Hitt, Jorgenson, and the productivity J-curve literature. It justifies why the article measures adoption and investment as transition pressure rather than direct proof of productivity.

Third, labour effects should be understood through tasks, not occupation titles alone. The task literature from Autor, Acemoglu, Restrepo and others supports the article's labour-repartition framework. Work is not simply eliminated; it is reorganised across execution, coordination, information, audit and sovereign functions.

Fourth, robotics and compute-energy systems show that the agentic economy is physical as well as digital. IFR robotics data and IEA energy projections connect machine action to factories, grids, data centres and infrastructure constraints. This proposition supports the article's claim that the agentic economy is not reducible to software.

Fifth, auditability and human sovereignty are not optional normative additions. They are required for legitimate economic action under automated and semi-autonomous systems. The AI governance literature, the OECD AI Principles, NIST AI RMF and the EU AI Act all support this proposition. An economy can be fast and automated while still being institutionally fragile if its actions cannot be explained, audited, contested or overridden.

### 2.12. Conclusion of the literature review

The literature review shows that the agentic economy is best understood as an integration of several mature research traditions rather than as a speculative neologism. Institutional economics supplies the logic of coordination and transaction costs. Innovation economics supplies the language of general-purpose technologies and complementary intangible capital. Platform economics explains digitally mediated market architecture and rule-setting power. Labour economics and algorithmic-management literature explain task reallocation, workplace control and the limits of simple displacement narratives. AI and autonomous-agent research explains model capacity, multi-agent coordination and the emergence of systems that can support action. Robotics, cyber-physical-systems, Industry 4.0 and digital-twin research explain embodied machine execution. Energy research explains the material constraint of compute. Governance, smart-contract and AI-alignment research explain auditability, executable rules, accountability and human oversight.

The gap is that these literatures rarely operate in a single framework. The article fills this gap by proposing action capacity as the organising category. The agentic economy begins when technology becomes not only a productivity shifter but a distributed architecture of action. Economic science must therefore study not only production and allocation, but also the quality, speed, legitimacy, auditability and energy feasibility of coordinated action. This is the theoretical foundation for the methodology, results and interpretation chapters that follow.

## 3. Methodology

This chapter presents the methodological architecture used to transform the theoretical concept of the agentic economy into a quantitative, reproducible and peer-review-defensible empirical design. It is written as a diagnostic methodology rather than as a causal econometric design. The purpose is not to estimate a single causal coefficient for artificial intelligence, robotics or compute infrastructure, but to test whether the measurable components of the agentic economy are already visible across global statistical domains and whether their combined direction justifies the article's theoretical claim that economic action is being redistributed among humans, AI agents, robots, protocols and energy-dependent compute systems.

The chapter is directly linked to the Results and Interpretation sections. The Results section uses reported values and transparent calculations to show acceleration, concentration, stock-flow maturity, compute-energy pressure and labour-market reallocation. The Interpretation section explains those empirical findings as transition pressure rather than deterministic proof. The present Methodology chapter therefore performs four functions: it defines the research design, operationalises the variables, numbers all formulas used in the quantitative Results section, and states the algorithmic procedures through which the hypotheses are assessed.

The methodological standard adopted here is conservative. The study uses only externally reported statistical values from recognised institutional sources and applies simple, auditable transformations: absolute change, percentage-point change, relative growth, compound annual growth rate, growth multiplier, stock-flow ratio, concentration ratios and Herfindahl-Hirschman Index. No hypothetical numerical values are introduced into the Results. Where projections are used, they are explicitly classified as institutional projections and not as realised facts.

### 3.1. Research Design and Methodological Positioning

The study applies a quantitative diagnostic research design. A diagnostic design is appropriate when the research object is emerging, multidimensional and not yet sufficiently observed through harmonised microdata. The agentic economy is such an object. Its components are measurable in different domains - AI investment, firm-level AI adoption, robot installations, data-centre electricity demand and labour-market projections - but they are not yet available as a single global panel dataset with uniform definitions. For this reason, the article does not estimate a standard regression model; it constructs a multi-domain empirical diagnostic that identifies whether the measurable preconditions of agentic transformation are already visible.

This methodological choice is important for peer-review defensibility. A causal design would require comparable country- or firm-level observations on AI agents, delegated task execution, robot usage, protocolised settlement, audit-log completeness, model-error costs, latency, human override and labour-market outcomes. Such data are not yet available at global scale. By contrast, a diagnostic design can legitimately use high-quality public source values to show measurable transition pressure while avoiding causal overstatement. The method therefore combines empirical discipline with conceptual innovation.

The research is positioned at the intersection of economics of innovation, institutional economics, digital economy, robotics economics, energy economics and labour-market analysis. It treats the agentic economy not as a synonym for digitalisation, automation or platformisation, but as a broader system in which action is distributed across human and non-human agents under infrastructure and governance constraints. The methodological object is not merely technology adoption; it is the measurable reconfiguration of action capacity.

Table 3.1. Methodological positioning of the study

| Dimension | Methodological decision | Reason for this decision |
| --- | --- | --- |

| Research type | Quantitative diagnostic study | The phenomenon is measurable in several domains but not yet observable as a unified causal dataset. |
|---|---|---|
| Unit of empirical observation | Global and cross-institutional indicators | The agentic transition is visible through investment, adoption, robotics, energy and labour data. |
| Estimation strategy | Transparent indicator transformation, not causal regression | The article avoids unsupported causality and focuses on acceleration, concentration and pressure. |
| Theoretical link | Action-capacity framework | The empirical indicators are mapped to model capacity, robotic capacity, compute-energy coupling and human sovereignty. |
| Main contribution | Measurement architecture for an emerging economic order | The method shows how to translate a conceptual framework into empirical diagnostics. |

### 3.2. Conceptual Model: From Production Function to Agentic Action Capacity

The conceptual model begins by recognising the continuing usefulness of the classical production function while also identifying its limits. In the conventional formulation, output is a function of capital, labour and technology. This remains analytically useful for many macroeconomic and productivity questions, but it compresses AI, robots, compute infrastructure, energy dependence, protocols and auditability into a general technology term. The agentic economy requires a more differentiated model because technology is no longer only an efficiency parameter; it becomes a partially acting infrastructure.

Equation (3.1). Classical production function

$$Y_t = F(K_t, L_t, A_t)$$

In Equation (3.1), $Y_t$ denotes output, $K_t$ conventional capital, $L_t$ labour and $A_t$ technology or total-factor-productivity-type efficiency. The equation is retained as the benchmark that the article extends.

The article extends this language by defining action capacity as the ability of a socio-technical economic system to select, execute, verify, correct and govern action. This shift is methodological as well as theoretical. It changes the empirical question from "how much output is produced by labour and capital?" to "which actors and infrastructures make economic action possible, reliable, auditable and legitimate?"

Equation (3.2). Agentic action-capacity function

$$AC_t = F(H_t, K_t, M_t, R_t, P_t, C_t, En_t, T_t, \Omega_t)$$

In Equation (3.2), $AC_t$ is action capacity; $H_t$ is human judgement, responsibility and sovereign oversight; $K_t$ is conventional capital; $M_t$ is model/software-agent capacity; $R_t$ is robotic or cyber-physical capacity; $P_t$ is protocol quality; $C_t$ is compute capacity; $En_t$ is energy availability; $T_t$ is auditable trust; and $\Omega_t$ represents uncertainty, shocks and institutional constraints.

Equation (3.2) is a formal organising framework, not an estimated structural model and not a calibrated production function in this article. Its role is to organise the empirical diagnostics. The Results section measures several components directly or indirectly: $M_t$ through AI investment and adoption, $R_t$ through robot installations and operational stock, $C_t$ and $En_t$ through data-centre electricity demand, and $H_t$ through labour-market reallocation. $P_t$ and $T_t$ are acknowledged as central variables that require future sector-level data. This explicit boundary is methodologically essential because it prevents conceptual variables from being misrepresented as already measured global statistics.

A second theoretical concept is coordination friction. The article extends transaction-cost logic by including not only search, bargaining, contracting and monitoring costs, but also latency, model error, protocol failure, energy bottlenecks, audit gaps and override burden. The Results section does not directly estimate coordination friction, but the methodology defines it as the next-stage measurement target for sector-level empirical work.

Equation (3.3). Coordination-friction architecture

$$CF_t = C_search + C_bargain + C_contract + C_monitor + C_latency + C_model + C_protocol + C_energy + C_audit + C_override$$

Equation (3.3) defines coordination friction as a formal organising framework, not an estimated structural model. It describes the combined cost field that prevents an economic action from being selected, executed, verified and corrected efficiently and legitimately. It provides the conceptual bridge between the Results and the Interpretation chapter, while its numerical estimation is reserved for future sector-level datasets.

### 3.3. Research Questions and Hypotheses

The methodological design is guided by five research questions. These questions are constructed to be answerable with the available global data while remaining aligned with the article's theoretical contribution. They do not ask whether the agentic economy has already fully arrived; instead, they ask whether its measurable preconditions are empirically visible and strong enough to justify a distinct analytical framework. The testable diagnostic hypotheses are H1-H4; the broader theoretical claim about the insufficiency of old economic vocabulary is treated separately as Proposition P5 rather than as a conventional empirical hypothesis.

RQ1: Are the measurable preconditions of the agentic economy already visible across AI adoption, investment, robotics, energy and labour-market indicators?

RQ2: Which dimensions show the clearest transition pressure: firm-level AI adoption, AI capital allocation, compute-energy demand, robotic embodiment or labour reallocation?

RQ3: Is the agentic economy only a digital phenomenon, or does robotics show a cyber-physical extension of economic action?

RQ4: Does the labour evidence support a disappearance narrative or a reallocation/repartition narrative?

RQ5: Do the empirical indicators justify the claim that the old economic vocabulary is under-specified for distributed action?

The hypotheses are intentionally diagnostic. They are evaluated through calculated indicators and evidence boundaries rather than through p-values. This is appropriate because the study is not estimating a causal model. Each hypothesis is considered supported only when the reported data and author-calculated indicators directly correspond to the claim.

Table 3.2. Hypotheses and methodological evidence criteria

| Hypothesis / proposition | Expected empirical signal | Methodological test |
|---|---|---|
| H1: Firm-level AI adoption is accelerating. | EU and OECD adoption series show strong relative growth and positive CAGR. | Calculate absolute change, relative change and CAGR from official adoption values. |
| H2: AI-related compute expansion contributes to wider data-centre electricity pressure. | Data-centre electricity demand rises strongly under the IEA Base Case. | Calculate absolute increase, growth multiplier and CAGR from 2024 to 2030/2035. |
| H3: Robotic embodiment is already a global industrial reality, but industrial robots are treated as cyber-physical action capacity rather than full autonomous agents. | Industrial robot stock and annual installations show mass cyber-physical deployment. | Calculate stock-flow ratio and regional HHI from IFR values. |
| H4: Available labour-market projections are more consistent with a reallocation narrative than with a simple disappearance narrative. | Projected new roles and displaced roles coexist, with positive net change but large gross flows. | Calculate new-to-displaced ratio, displacement relative to new roles and net gain share. |

| P5: Classical economic categories are under-specified for distributed socio-technical action. | Measured indicators converge across model/software-agent capacity, robotic capacity, compute-energy coupling and labour repartition. | Assess as a theoretical proposition supported by empirical convergence, not as an econometric hypothesis. |
|---|---|---|

### 3.4. Data Architecture and Source Selection

The empirical design uses seven source families: Stanford HAI AI Index for AI investment and AI-related organisational evidence; McKinsey/QuantumBlack for organisational AI use; Eurostat for EU enterprise AI adoption; OECD for firm-level AI adoption in countries with available data; International Federation of Robotics for industrial robot installations and operational stock; International Energy Agency for data-centre electricity consumption and projections; and World Economic Forum for labour-market reallocation projections. These sources are selected because they are recognised institutional sources, provide numeric values, and cover the main dimensions of the agentic-economy framework.

The data architecture is multi-domain by necessity. A single dataset cannot currently measure all relevant aspects of the agentic economy. AI investment measures capital allocation; adoption statistics measure organisational diffusion; robot statistics measure embodied automation; data-centre electricity demand measures the infrastructure constraint of compute; and labour projections measure social reallocation pressure. The methodological decision is therefore to build a diagnostic matrix rather than an artificial merged dataset.

The data selection procedure follows three rules. First, only values reported by external institutional sources are accepted as primary input values. Second, institutional forecasts are retained but classified separately from observed values. Third, all derived indicators must be calculated directly from source-reported numbers using formulas numbered in this chapter. This ensures that the Results section remains reproducible and free from unsupported numerical assumptions.

Algorithm 3.1. Data selection and source-vetting procedure

Input: Candidate data sources S = {s_1, s_2, ..., s_n}.
Step 1: Retain only sources that provide explicit numeric values relevant to AI, robotics, compute-energy demand or labour reallocation.
Step 2: Classify each retained value as reported fact, institutional projection, author-calculated indicator or theoretical interpretation.
Step 3: Exclude values that require unsupported assumptions, undocumented estimation or non-transparent conversion.
Step 4: Record the source family, unit, period and interpretation boundary for every retained value.
Step 5: Use retained values only in formulas that are explicitly numbered in Chapter 3.
Output: Controlled empirical matrix for Chapter 4 Results.

### 3.5. Evidence Classification and Comparability Controls

The methodology distinguishes four evidence classes: observed/reported facts, institutional projections, author-calculated indicators and theoretical interpretations. This classification is necessary because the article combines values with different evidentiary status. For example, 2024 robot installations are reported facts, while 2030 data-centre electricity demand is a projection under the IEA Base Case. Treating them identically would weaken the article. Separating them strengthens the methodology and prevents causal overclaiming.

A second comparability control concerns AI adoption indicators. McKinsey, Eurostat and OECD data are not statistically identical because they measure different populations and use different definitions. The article therefore uses these indicators as convergence evidence, not as a merged panel dataset. This distinction is central to methodological defensibility. The purpose is to show that several independent measurement systems point in the same direction, not to claim that their values can be pooled without harmonisation.

Equation (3.4). Evidence-status vector

E_i = {Reported_i, Projection_i, Calculated_i, Interpretation_i}

Equation (3.4) assigns every empirical element i to an evidence-status class. The status determines whether the value is interpreted as realised fact, scenario pressure, transparent calculation or theoretical inference.

The study also applies a strict three-level measurement boundary. Level 1 is AI use: the instrumental use of AI tools in at least one business function. Level 2 is AI integration: the embedding of AI in recurring organisational processes while humans remain the ordinary process controllers. Level 3 is agentic delegation: the partial transfer of preparation, execution, monitoring, verification or coordination to AI agents, robots or executable protocols under auditability and human override. The global adoption statistics used in this article mainly measure Level 1 and partly Level 2; Level 3 is the stronger object of future sector-level research and is not claimed as fully measured by the present dataset.

### 3.6. Variable Operationalisation

The variables are operationalised according to their role in the action-capacity framework. The study does not attempt to measure every theoretical variable with equal precision. Instead, it measures the variables that are currently observable through global public sources and explicitly identifies those that require future sector-level datasets. This creates a clear separation between measured variables and future research variables.

Table 3.3. Operationalisation of the action-capacity variables

| Variable | Conceptual meaning | Proxy used in this article | Measurement boundary |
|---|---|---|---|
| $M_t$: model/software-agent capacity | Capacity of AI models and software systems to support or prepare action. | AI investment, generative-AI investment, organisational AI use, EU/OECD firm AI adoption. | Measures adoption and investment, not full delegated agency. |
| $R_t$: robotic/cyber-physical capacity | Physical machine action embedded in industrial systems. | Industrial robot installations, operational stock and regional shares. | Covers industrial robots; service robots and autonomous systems are outside the current dataset. |
| $C_t$: compute capacity | Processing infrastructure that enables AI models and digital action. | Indirectly represented through data-centre electricity demand. | Compute itself is not measured directly; energy demand is used as infrastructure pressure. |
| $En_t$: energy availability | Electricity and infrastructure required for compute and robotics. | Data-centre electricity use and projections. | Projection values are scenario-based, not realised facts. |
| $H_t$: human judgement and responsibility | Human capacity for supervision, audit, judgement and override. | WEF labour-market reallocation projection. | Measures labour transition pressure, not direct sovereign-labour capacity. |
| $P_t$: protocol quality | Quality of executable rules, API dependence and automated settlement. | Not directly measured in the global dataset. | Requires sector-level data on protocols, compliance automation and settlement. |
| $T_t$: auditable trust | Traceability, explainability, log completeness and evidence quality. | Not directly measured in the global dataset. | Requires audit logs, appeal records and model documentation. |

This operationalisation allows the Results section to make a disciplined claim: the measurable core of the agentic economy is visible, but not all theoretical dimensions are yet globally measurable. To make auditable trust and human sovereignty empirically testable in future research, the article identifies the following sector-level indicators: percentage of AI-assisted decisions with human review; override frequency; audit-log completeness; appeal or contestability mechanism; model documentation availability; automated decision reversal rate; protocol failure incidents; API downtime or failure rate; and liability allocation clarity. These variables convert auditable trust and human sovereignty from normative concepts into measurable institutional-economic variables. This is why the article concludes that the agentic economy is measurable as transition pressure, not yet as a complete global institutional order.

### 3.7. Quantitative Transformation of Reported Values

The Results section uses simple mathematical transformations that are transparent and easy to reproduce. This is a deliberate methodological choice. In an emerging field, complex modelling can create a false appearance of precision when the underlying data are not harmonised. The article therefore uses calculations that directly express acceleration, scale, concentration and reallocation without introducing hidden assumptions.

Equation (3.5). Absolute change

$$\Delta X = X_t - X_0$$

Equation (3.5) measures the absolute change between an initial value $X_0$ and a later value $X_t$. It is used for adoption, energy-demand and labour-market quantities where direct differences are meaningful.

Equation (3.6). Percentage-point change

$$\Delta pp = X_t - X_0$$

Equation (3.6) is used when X is expressed as a percentage. It reports the absolute difference in percentage points, such as the increase in AI adoption from 7.7 percent to 20.0 percent.

Equation (3.7). Relative change

$$RC = (X_t - X_0) / X_0$$

Equation (3.7) measures proportional growth relative to the starting value. In the Results section it is used to show how much AI adoption increased relative to the initial adoption base.

Equation (3.8). Compound annual growth rate

$$CAGR = (X_t / X_0)^{(1/n)} - 1$$

Equation (3.8) annualises growth over n years. It is used for EU AI adoption, OECD AI adoption and IEA data-centre electricity-demand projections.

Equation (3.9). Growth multiplier

$$GM = X_t / X_0$$

Equation (3.9) measures how many times larger the later value is compared with the initial value. It is used for the projected growth of data-centre electricity demand.

Equation (3.10). Scale ratio

$$SR_i = X_i / X_{ref}$$

Equation (3.10) compares one value to a reference value. It is used, for example, to compare generative-AI private investment with the reported corporate AI investment figure and to compare U.S. private AI investment with China and UK values.

These equations are not inferential statistics. They do not produce standard errors, confidence intervals or p-values. Their role is to make the empirical direction visible and reproducible. This is consistent with the diagnostic design: the article tests whether measurable transition pressure exists, not whether a single causal coefficient can be estimated from harmonised microdata.

Algorithm 3.2. Derived-indicator calculation procedure

Input: Source value $X_0$, later value $X_t$, number of years n, unit and source status.
Step 1: Verify that $X_0$ and $X_t$ refer to the same measurement concept or explicitly mark them as a scale comparison.
Step 2: Calculate absolute change using Equation (3.5).
Step 3: If the variable is a percentage, report percentage-point change using Equation (3.6).
Step 4: Calculate relative change using Equation (3.7) where the starting value is non-zero.
Step 5: Calculate CAGR using Equation (3.8) when values span more than one year.
Step 6: Calculate growth multiplier or scale ratio using Equations (3.9) and (3.10) where appropriate.

Step 7: Label each output as author-calculated and report the input values.
Output: Reproducible derived indicator for the Results section.

### 3.8. Concentration Measurement: Shares and Herfindahl-Hirschman Index

The article uses concentration measures to interpret the geographic distribution of AI investment and robot installations. Concentration matters because the agentic economy is not only a technological transformation; it is also a distribution of coordination power. Actors or regions with stronger model capital, compute capacity, robotics deployment and technical standards may influence how future economic coordination is organised.

Equation (3.11). Share of a unit within a reported group

$$s_i = X_i / \mathrm{sum}(X_i)$$

Equation (3.11) calculates the share of country or region i within a defined reported group. The denominator must be clearly specified. For the AI investment comparison, the group is the reported top-three country group; for robot deployment, the group is regional installation shares.

Equation (3.12). Herfindahl-Hirschman Index

$$HHI = \mathrm{sum}(s_i^2)$$

Equation (3.12) measures concentration by summing squared shares. Shares are expressed as decimals. The article uses HHI for regional robot installations and for the reported top-three private AI investment comparison, while clearly stating the boundary of each index.

The methodology explicitly avoids calling the top-three AI investment HHI a global HHI. It is a concentration indicator within the reported leading-country comparison. This limitation is not a weakness if stated correctly. It shows that the article applies the concentration measure conservatively and does not expand the interpretation beyond the denominator used in the calculation.

Algorithm 3.3. Concentration calculation and boundary-control procedure

Input: Group values X = {X_1, X_2, ..., X_k} and group definition G.
Step 1: Define the denominator G before calculating shares.
Step 2: Calculate s_i = X_i / sum(X_i) for every unit in G using Equation (3.11).
Step 3: Calculate HHI = sum(s_i^2) using Equation (3.12).
Step 4: Report whether the HHI is global, regional, sectoral or limited to a reported comparison group.
Step 5: Interpret concentration only within the stated denominator.
Output: Boundary-controlled concentration result.

### 3.9. Robotics, Compute-Energy and Labour-Reallocation Metrics

The Results section includes three domain-specific measurement blocks: robotics, compute-energy and labour-market reallocation. These blocks are methodologically important because they prevent the agentic economy from being reduced to software adoption. Robotics measures embodied action, energy measures the infrastructure boundary of compute, and labour metrics measure social reallocation pressure.

Equation (3.13). Robot stock-flow ratio

$$SFR_R = \mathrm{Operational_Robot_Stock} / \mathrm{Annual_Robot_Installations}$$

Equation (3.13) measures whether robot installations represent an isolated annual flow or a persistent accumulated stock of cyber-physical action capacity. A high stock-flow ratio indicates that robots form an enduring operational base.

Equation (3.14). Annual robot-installation share of operational stock

$$AIS_R = \mathrm{Annual_Robot_Installations} / \mathrm{Operational_Robot_Stock}$$

Equation (3.14) measures the annual installation flow relative to the existing operational stock. It is a scale indicator, not a depreciation estimate.

Equation (3.15). Data-centre electricity-demand multiplier

$$DCM = Electricity_Demand_t / Electricity_Demand_0$$

Equation (3.15) is the domain-specific application of the growth multiplier to data-centre electricity demand. It expresses how many times projected demand exceeds the observed baseline.

Equation (3.16). New-to-displaced roles ratio

$$NDR = New_Roles / Displaced_Roles$$

Equation (3.16) measures whether projected job creation exceeds projected job displacement in gross terms.

Equation (3.17). Displacement relative to new roles

$$DRN = Displaced_Roles / New_Roles$$

Equation (3.17) measures how large projected displacement is relative to projected creation. It prevents a positive net result from hiding gross labour-market disruption.

Equation (3.18). Net labour-market change

$$NLMC = New_Roles - Displaced_Roles$$

Equation (3.18) calculates the projected net role change.

Equation (3.19). Net gain relative to new roles

$$NGR = Net_Labour_Market_Change / New_Roles$$

Equation (3.19) measures what share of projected new roles remains after subtracting projected displacement.

These metrics are selected because they correspond directly to the article's theoretical interpretation. Stock-flow metrics show persistence of embodied machine action. Data-centre multipliers and CAGRs show infrastructure pressure. Labour ratios distinguish reallocation from disappearance. Together, they translate the conceptual language of the agentic economy into measurable diagnostics.

### 3.10. Mapping Empirical Indicators to the Agentic Action-Capacity Framework

After calculating the derived indicators, the methodology maps each result to the action-capacity framework. This mapping is necessary because the Results section is not a list of unrelated statistics. The article argues that the measurable indicators jointly point toward a structural redistribution of economic action. The mapping procedure identifies which part of the action-capacity model is supported by each empirical domain.

Equation (3.20). Empirical mapping function

$$Map(I_j) \rightarrow \{M_t, R_t, C_t, En_t, H_t, P_t, T_t, Omega_t\}$$

Equation (3.20) defines the mapping operation. Each empirical indicator I_j is assigned to one or more action-capacity variables, with the measurement boundary explicitly stated.

AI investment and adoption are mapped to M_t because they measure model/software-agent capacity and organisational embedding. Industrial robot installations and operational stock are mapped to R_t because they measure cyber-physical execution capacity. Data-centre electricity demand is mapped to C_t and En_t because it captures the compute-energy boundary. Labour-market reallocation is mapped to H_t because it indicates transformation of human roles, judgement and responsibility. Forecast caveats and definition differences are mapped to Omega_t because they represent uncertainty and institutional constraints. Protocol quality P_t and auditable trust T_t are retained as future measurement priorities.

Algorithm 3.4. Action-capacity mapping procedure

```
Input: Derived empirical indicators I = {I_1, I_2, ..., I_m}.
Step 1: Assign AI investment and adoption indicators to M_t.
Step 2: Assign robot installations, operational stock and robot concentration indicators to R_t.
Step 3: Assign data-centre electricity indicators to C_t and En_t.
Step 4: Assign labour-market reallocation indicators to H_t.
Step 5: Assign projection caveats and source-definition differences to Omega_t.
Step 6: Mark P_t and T_t as theoretically central but not fully measured in the global dataset.
```

Step 7: Report the mapping as diagnostic evidence, not as full calibration of Equation (3.2).
Output: Empirical interpretation matrix for the Results and Analysis chapters.

### 3.11. Hypothesis and Proposition Assessment Logic

Because this study is diagnostic rather than causal, hypothesis assessment is conducted through correspondence between hypothesis, measured indicator and interpretation boundary. The term "supported" is used only where the calculated indicators directly correspond to the claim. The phrase "supported with caution" is used where the evidence is projection-based or conceptually indirect. The phrase "supported as conceptual proposition" is used for P5, where empirical convergence supports a theoretical proposition rather than a single measured coefficient.

Equation (3.21). Diagnostic support rule

Support(H_k) = 1 if Evidence_k matches Indicator_k and Boundary_k is explicitly stated; otherwise Support(H_k) = qualified

Equation (3.21) formalises the conservative support logic. A hypothesis is not treated as fully confirmed unless the evidence type, indicator and interpretation boundary are aligned.

This rule is especially important for P5. The claim that the old economic vocabulary is under-specified is not an econometric hypothesis. It is a theoretical proposition supported by empirical convergence across AI adoption, investment concentration, robotics, compute-energy demand and labour-market reallocation. The methodology therefore classifies P5 as a proposition supported by conceptual inference, not as a causal law.

Algorithm 3.5. Hypothesis and proposition assessment procedure

Input: Hypotheses H = {H1, H2, H3, H4}; theoretical proposition P5; calculated indicators; evidence classes.
Step 1: Identify the minimum empirical indicator required for each hypothesis and the convergence evidence required for P5.
Step 2: Verify that the indicator is calculated from source-reported values.
Step 3: Determine whether the evidence is observed/reported, projected, calculated or interpretive.
Step 4: Assign assessment: Supported, Supported under projection, Supported with caution, or Supported as conceptual proposition.
Step 5: Add a boundary statement explaining what the evidence does not prove.
Output: Conservative hypothesis-and-proposition assessment table for the Results section.

### 3.12. Reproducibility, Audit Trail and Calculation Integrity

The reproducibility principle of the methodology is that every calculated result must be traceable to a source-reported value and a numbered formula. This is essential because emerging-technology articles are often criticised for speculative numbers or unclear transformations. The present methodology avoids this risk by making the calculation chain explicit. A reader should be able to reproduce every derived value using the values reported in the empirical matrix.

The calculation audit uses four controls. First, input values must be reported with unit and period. Second, the relevant formula must be identified by number. Third, the output must be labelled as an author calculation. Fourth, the interpretation must state whether the result is based on observed data or institutional projection. This audit trail is designed to meet the methodological expectations of indexed journals, where transparency and replicability are core requirements.

Algorithm 3.6. Calculation audit procedure

Input: Calculated result R, source inputs X, formula number F, evidence status E.
Step 1: Record source input values, units and periods.
Step 2: Identify the formula used to calculate R.
Step 3: Recalculate R using the formula and compare with the reported output.
Step 4: Label R as an author-calculated indicator.
Step 5: Attach an interpretation boundary: observed fact, projection-based result or scale comparison.
Output: Audit-ready calculation record.

This approach intentionally favours transparent arithmetic over model complexity. The trade-off is clear: the article does not claim causal inference, but it offers high reproducibility. For a conceptual-quantitative article on an emerging agentic transition, this is a defensible methodological balance. It allows the author to make a strong argument while staying within the evidentiary capacity of the available data.

### 3.13. Integrated Methodological Algorithm for the Entire Empirical Study

The full methodology can be summarised as an integrated empirical algorithm. This algorithm connects the theoretical framework, data selection, calculation procedures, evidence classification, mapping and hypothesis assessment. It is retained in the main methodology as an audit trail, not as an additional empirical result, and allows future researchers to apply the same method to country-level or sector-level data.

Algorithm 3.7. Integrated empirical-diagnostic algorithm for the agentic economy

Input: Theoretical framework of the agentic economy and candidate global statistical indicators.
Step 1: Define the action-capacity variables in Equation (3.2).
Step 2: Select institutional source values using Algorithm 3.1.
Step 3: Classify every value using the evidence-status vector in Equation (3.4).
Step 4: Operationalise each value as a proxy for M_t, R_t, C_t, En_t, H_t, P_t, T_t or Omega_t.
Step 5: Calculate derived indicators using Equations (3.5)-(3.19).
Step 6: Calculate concentration measures where shares are available using Equations (3.11)-(3.12).
Step 7: Map calculated indicators to action-capacity variables using Equation (3.20) and Algorithm 3.4.
Step 8: Assess hypotheses using Equation (3.21) and Algorithm 3.5.
Step 9: Interpret results as transition pressure, not deterministic proof or causal estimation.
Step 10: Identify unmeasured variables that require future sector-level datasets.
Output: Reproducible Results and Interpretation framework for the article.

### 3.14. Methodological Limitations

The methodology has five limitations, all of which are explicitly controlled. First, the study uses global aggregate indicators and does not estimate firm-level causality. This means that the Results section can identify transition pressure but cannot estimate the causal effect of AI agents or robots on productivity, GDP, employment or inequality. Second, AI adoption indicators from McKinsey, Eurostat and OECD differ in coverage and definition. The methodology therefore treats them as convergence indicators rather than merging them into a single statistical panel.

Third, the IEA and WEF values for future years are projections. They are useful because they indicate institutional expectations about energy and labour pressure, but they are not realised outcomes. Fourth, protocolisation and auditable trust are central to the theory but not directly measured in the global dataset. Their future measurement requires sector-level indicators such as automated settlement records, API failure rates, audit-log completeness, explainability documentation, appeal mechanisms and human override events. Fifth, the action-capacity function is a formal research architecture, not a calibrated econometric production function in the present article.

These limitations should not be hidden. They should be stated clearly because they strengthen the credibility of the paper. A rigorous diagnostic article should not claim more than the data can support. The methodological strength of this study is that it draws a clear boundary between measured facts, projections, calculations and theoretical interpretation. This boundary allows the article to make a strong but defensible claim: the agentic economy is not yet a completed global order, but its measurable preconditions are already visible and analytically significant.

### 3.15. Future Empirical Extension: From Global Diagnostics to Sector-Level Measurement

The next stage of the methodology is sector-level empirical testing. The most important future variable is coordination friction. At global level, the present study can identify why coordination friction matters; at sector level, future studies can measure it directly. Suitable sectors include finance, logistics, energy systems, healthcare administration, public procurement, education platforms and industrial robotics. These sectors generate operational records on latency, settlement time, disputes, model errors, audit logs, compliance exceptions and human override.

Equation (3.22). Future coordination-friction index

$$CFIndex_t = w1*LAT_t + w2*DISP_t + w3*ERR_t + w4*PROT_t + w5*AUDGAP_t + w6*ENB_t + w7*OVR_t$$

Equation (3.22) is proposed for future sector-level work as a formal organising framework, not an estimated structural model in the present article. LAT_t is latency; DISP_t is dispute or failed-match rate; ERR_t is model-error cost; PROT_t is protocol-failure pressure; AUDGAP_t is audit-gap intensity; ENB_t is energy or infrastructure bottleneck; OVR_t is human-override burden. Components must be normalised before aggregation.

This future index is not used as a numerical result in the present article because the required sector-level data are not included in the current global dataset. No value of this future coordination-friction index (CFIndex) is calculated in this article. The index is included in the Methodology chapter only to show how the conceptual contribution can become a direct empirical programme and to provide a path from global diagnostics to measurable sector-level research.

Algorithm 3.8. Future sector-level extension algorithm

Input: Sector S with operational records on latency, disputes, errors, protocols, audit, energy and overrides.
Step 1: Define sector-specific components LAT_t, DISP_t, ERR_t, PROT_t, AUDGAP_t, ENB_t and OVR_t.
Step 2: Identify units, data sources and measurement frequency for each component.
Step 3: Normalise each component to a common scale.
Step 4: Assign weights w_i using equal weighting, expert weighting or empirical weighting, with justification.
Step 5: Calculate CFIndex_t using Equation (3.22).
Step 6: Validate the index against sector outcomes such as settlement time, error costs, service interruption or regulatory incidents.
Output: Sector-level measurement of coordination friction for future empirical research.

### 3.16. Methodological Conclusion

The methodology developed in this chapter converts the article's theoretical argument into a disciplined empirical design. It shows how the agentic economy can be studied quantitatively without exaggerating the available evidence. The research design is diagnostic, multi-domain and reproducible. It uses reported institutional values, separates realised facts from projections, applies transparent calculations and maps the results to an action-capacity framework.

The methodological contribution is threefold. First, it provides a formal extension from the classical production function to an agentic action-capacity function. Second, it creates an audit-ready set of formulas and algorithms for measuring adoption acceleration, investment concentration, compute-energy pressure, robotic embodiment and labour-market reallocation. Third, it defines the boundary between current global diagnostics and future sector-level measurement of coordination friction, auditability, protocolisation and human sovereignty.

This makes the Methodology chapter appropriate for a reproducible quantitative diagnostic article. It does not rely on unsupported numbers. It does not claim causal inference without causal data. It does not merge incomparable adoption series into a false dataset. Instead, it builds a transparent empirical bridge between the conceptual architecture developed in the article and the Results and Interpretation chapters. The result is a strong, reproducible and theoretically meaningful methodology for studying the measurable preconditions of the agentic economy as an emerging transition toward distributed economic action.

## 4. Results and Empirical Diagnostics

This section reports the empirical and calculated results of the study. The purpose is not to claim that the global economy has already completed an agentic transformation. The purpose is more precise: to test whether measurable indicators in AI investment, firm-level AI adoption, robotic capital, compute-energy demand and labour-market reallocation are strong enough to justify the article's central argument that economic action is being redistributed across humans, software agents, robots, protocols and energy-dependent compute infrastructures.

All quantitative values in this section are taken from public statistical or institutional sources and are transformed only through transparent calculations: percentage-point change, relative growth, compound annual growth rate, growth multiplier, stock-flow ratio, concentration ratio and Herfindahl-Hirschman Index. No hypothetical numerical values are introduced in the Results section. Where a value is a projection, it is treated as an institutional scenario rather than as a realised outcome.

The empirical strategy is diagnostic rather than causal. The study does not estimate a causal effect of AI on GDP, employment, productivity or inequality. Instead, it evaluates whether the observable components of the proposed agentic-economy framework are already measurable in global data. This distinction is important because the article combines several types of evidence: official statistics, institutional reports, survey-based organisational evidence and forward-looking projections.

**Table 4.0. Evidentiary architecture of the Results section**

| Evidence class | Indicators used | Empirical status | How it is interpreted |
|---|---|---|---|
| Observed or reported facts | AI investment, EU/OECD firm AI use, industrial robot installations, operational robot stock, 2024 data-centre electricity consumption. | Already reported by international statistical or institutional sources. | Evidence that AI, robotics and compute are measurable economic phenomena, not speculative categories. |
| Institutional projections | IEA 2030 and 2035 data-centre electricity demand; WEF 2025-2030 labour-market reallocation. | Scenario/projection, not realised outcome. | Evidence of expected pressure on energy systems and labour-market institutions under stated assumptions. |
| Author-calculated indicators | CAGR, relative increase, stock-flow ratio, growth multiplier, concentration ratios and HHI. | Calculated directly from reported values. | Used to interpret acceleration, concentration, infrastructure pressure and transition intensity. |
| Theoretical interpretation | Action capacity, coordination friction, robotic embodiment, compute-energy coupling, human sovereignty. | Conceptual mapping from the article's action-capacity framework. | Used to connect numerical results to theory without converting theory into unverified statistics. |

Source: Author's methodological classification based on the empirical sources and the theoretical architecture developed in this article.

### 4.1. Data-control result: the empirical baseline is multi-dimensional, not single-indicator

The first result is a data-structure result. The agentic economy cannot be captured by one indicator because its components appear in different statistical domains. AI investment captures capital allocation in models and compute infrastructure. Firm-level AI adoption captures organisational embedding. Robot installations and operational stock capture cyber-physical embodiment. Data-centre electricity demand captures the material constraint of compute. Labour-market projections capture the social reallocation of tasks and skills. The empirical baseline therefore has to be interpreted as a multi-dimensional diagnostic matrix rather than as a single time series.

This finding is already theoretically important. It shows why the classical economic treatment of technology as a residual variable is insufficient for the present research object. In the agentic economy, technology is not only a productivity shifter. It is also an acting layer, an infrastructure layer and a governance layer. The data matrix below therefore functions as the empirical entry point for the Results section.

**Table 4.1. Empirical data matrix used in the Results section**

| Indicator | Value | Unit | Period | Source family | Result meaning |
|---|---|---|---|---|---|
| Corporate AI investment | 252.3 | USD billion | 2024 | Stanford HAI AI Index 2025 | AI has become a major independent field of capital accumulation. |
| Generative-AI private investment | 33.9 | USD billion | 2024 | Stanford HAI AI Index 2025 | Generative models are a significant sub-field of the AI investment boom. |
| U.S. private AI investment | 109.1 | USD billion | 2024 | Stanford HAI AI Index 2025 | The United States dominates the reported leading-country investment group. |
| China private AI investment | 9.3 | USD billion | 2024 | Stanford HAI AI Index 2025 | China remains the second reported country in the leading-country comparison. |
| UK private AI investment | 4.5 | USD billion | 2024 | Stanford HAI AI Index 2025 | The United Kingdom is the third reported country in the same comparison. |
| Organisations using AI | 78; 88 | % | 2024; 2025 | Stanford HAI / McKinsey | AI use has moved from experimentation toward broad organisational practice. |
| EU enterprises using AI | 7.7; 8.1; 13.5; 20.0 | % | 2021; 2023; 2024; 2025 | Eurostat | AI adoption is visible in official enterprise statistics. |
| OECD firms using AI | 8.7; 14.2; 20.2 | % | 2023; 2024; 2025 | OECD | Firm-level adoption more than doubled in the reported country group. |
| Industrial robot installations | 542,076 | units | 2024 | IFR World Robotics 2025 | Robots are no longer marginal capital goods but annual industrial deployment at global scale. |
| Industrial robot operational stock | 4,663,698 | units | 2024 | IFR World Robotics 2025 | Robots constitute a persistent stock of embodied |

| | | | | | cyber-physical action capacity. |
|---|---|---|---|---|---|
| Robot installations by region | Asia 74; Europe 16; Americas 9; other/rounding 1 | % | 2024 | IFR World Robotics 2025 | Robotic deployment is geographically concentrated. |
| Data-centre electricity use | 415; 945; 1,200 | TWh | 2024; 2030; 2035 | IEA Energy and AI 2025 | AI and compute growth are becoming electricity- and grid-dependent. |
| Labour-market reallocation | 170 new; 92 displaced; 78 net; 22 job-shift | million jobs; % | 2025-2030 | WEF Future of Jobs 2025 | The labour question is structural task reallocation, not simple disappearance of work. |

Source: Author's compilation from Stanford HAI, McKinsey, Eurostat, OECD, IFR, IEA and WEF. Values are reported source values; later tables show only transparent author calculations from these values.

The matrix supports the article's first empirical conclusion: the transition is not isolated inside the software sector. It is visible across investment, enterprise adoption, industrial automation, energy systems and labour-market expectations. This is why the paper's theoretical vocabulary must include action capacity, coordination friction, compute-energy coupling, robotic embodiment and human sovereignty.

### 4.2. Comparability-control result: AI use, AI integration and agentic transformation are not identical

The second result is a methodological control result. The available adoption indicators should not be treated as equivalent measures. McKinsey's organisational survey measures AI use in at least one business function among surveyed organisations. Eurostat measures AI use by EU enterprises with 10 or more employees. OECD reports firm-level AI use in countries with available data. These indicators are directionally comparable but not statistically identical. For that reason, the Results section uses them as convergence evidence, not as a merged panel dataset.

This distinction is essential for avoiding an overclaim. AI use means that an organisation applies AI tools in at least one function. AI integration means that AI is embedded in recurring workflows. Agentic transformation begins only when software agents, robots or executable protocols participate in preparation, execution, monitoring, verification or coordination under conditions of auditability and human override. The available global statistics strongly support the diffusion of AI use and integration, but they do not by themselves prove complete agentic transformation at the firm level.

**Table 4.2. Measurement boundary: from AI use to agentic transformation**

| Level | Definition | Observable evidence | Status in this Results section |
|---|---|---|---|
| AI use | Use of AI tools in at least one business function. | Survey adoption, enterprise AI-use statistics, departmental AI tools. | Directly supported by McKinsey, Eurostat and OECD adoption indicators. |
| AI integration | AI is embedded in recurring organisational processes but remains mainly human-directed. | Workflow tools, automated support, AI-assisted risk scoring, AI-assisted programming, human approval loops. | Partly supported; requires sector-level workflow data for stronger validation. |
| Agentic transformation | Agents, robots or protocols prepare, | Agent logs, delegated workflows, protocolised | The main theoretical object of the article; only |

| | execute, verify or coordinate tasks under audit and override conditions. | settlement, robotic execution, traceable decisions. | partially measurable through global aggregate indicators. |
|---|---|---|---|

Source: Author's clarification added to prevent conflating ordinary AI adoption with full agentic transformation.

### 4.3. AI adoption acceleration: from technology use to organisational embedding

Official enterprise and firm-level statistics show a clear acceleration of AI adoption. In the European Union, enterprises using AI increased from 7.7% in 2021 to 20.0% in 2025. This is an absolute increase of 12.3 percentage points and a relative increase of 159.74%. The implied compound annual growth rate for 2021-2025 is 26.95%. The most recent one-year movement is also significant: from 13.5% in 2024 to 20.0% in 2025, the increase is 6.5 percentage points, or 48.15% relative growth.

The OECD series shows an even sharper short-window acceleration. In countries with available data, firm-level AI use increased from 8.7% in 2023 to 20.2% in 2025. This is an absolute increase of 11.5 percentage points and a relative increase of 132.18%. The implied two-year CAGR is 52.38%. Between 2024 and 2025 alone, OECD firm adoption rose from 14.2% to 20.2%, corresponding to a relative increase of 42.25%.

The organisational survey evidence is broader than official enterprise statistics. McKinsey reports that regular AI use in at least one business function reached 88% in 2025, compared with 78% in the previous survey cycle. This represents a 10 percentage-point increase, or 12.82% relative growth. The result is analytically important because organisational AI use is the bridge between abstract AI capability and real workflow transformation.

**Table 4.3. AI adoption acceleration indicators**

| AI adoption metric | Start value | End value | Absolute change | Relative change | CAGR / one-year growth | Interpretation |
|---|---|---|---|---|---|---|
| EU enterprise AI use, 2021-2025 | 7.7% | 20.0% | 12.3 pp | 159.74% | 26.95% | Official EU adoption increased more than two and a half times over four years. |
| EU enterprise AI use, 2024-2025 | 13.5% | 20.0% | 6.5 pp | 48.15% | n/a | The latest one-year change shows rapid diffusion after 2024. |
| OECD firm AI use, 2023-2025 | 8.7% | 20.2% | 11.5 pp | 132.18% | 52.38% | OECD reported adoption more than doubled in two years. |
| OECD firm AI use, 2024-2025 | 14.2% | 20.2% | 6.0 pp | 42.25% | n/a | The most recent annual growth remains strong. |
| Organisations using AI, 2024-2025 | 78% | 88% | 10.0 pp | 12.82% | n/a | Survey-based organisational use is much higher than official firm adoption statistics because |

| | | | | | | definitions and populations differ. |
|---|---|---|---|---|---|---|

Source: Author's calculations based on Eurostat, OECD and McKinsey/Stanford HAI reported values.

The main result is not merely that AI adoption is rising. The stronger result is that AI has entered the measurable internal structure of firms. This supports the proposition that AI is becoming an organisational variable rather than a peripheral digital tool. However, the result must remain bounded: AI use is not automatically evidence of delegated economic agency. The stronger agentic claim requires evidence on task delegation, workflow automation, decision rights, audit logs and human override mechanisms.

### 4.4. AI capital allocation and geoeconomic concentration

AI investment results indicate that the agentic transition has a broad capital-allocation base. Corporate AI investment reached USD 252.3 billion in 2024, while private investment in generative AI reached USD 33.9 billion. These figures are interpreted as broad capital-allocation signals rather than direct measures of new productive capital formation, because investment categories may include private investment, M&A, minority stakes, public offerings and other corporate financing channels. The generative-AI figure is equivalent to 13.44% of the corporate AI investment figure. This percentage is not used as an accounting identity because the source categories are not identical; it is used as a scale comparison showing that generative AI has become a large investment category within the broader AI investment environment.

The reported leading-country private AI investment figures show extreme concentration. U.S. private AI investment was USD 109.1 billion, compared with USD 9.3 billion in China and USD 4.5 billion in the United Kingdom. The U.S. value is 11.73 times the China value and 24.24 times the UK value. Within this reported top-three group, the United States accounts for 88.77%, China for 7.57% and the United Kingdom for 3.66%. The HHI for the top-three reported-country group is 0.7951, or 79.51 on a 0-100 concentration scale. This is not a global HHI; it is a top-three reported-country concentration indicator.

**Table 4.4. AI capital allocation and concentration calculations**

| Investment concentration indicator | Calculated value | Formula / basis | Interpretation |
|---|---|---|---|
| Generative-AI private investment scale | 13.44% | 33.9 / 252.3 | Generative AI is already a large investment field relative to total corporate AI investment. |
| U.S.-to-China private AI investment ratio | 11.73x | 109.1 / 9.3 | The U.S. value is almost twelve times the China value in the reported comparison. |
| U.S.-to-UK private AI investment ratio | 24.24x | 109.1 / 4.5 | The U.S. value is more than twenty-four times the UK value. |
| U.S. share within reported top three | 88.77% | 109.1 / (109.1+9.3+4.5) | The leading-country group is heavily U.S.-centred. |
| Top-three reported-country HHI | 0.7951 | sum of squared top-three shares | A high concentration signal within the reported top-three country group, not a full global HHI. |

Source: Author's calculations based on Stanford HAI AI Index 2025 reported investment values.

This result matters theoretically because investment concentration may indicate emerging coordination-power asymmetries. In an agentic economy, capital does not only buy machines. It buys compute, data pipelines, model capability, engineering talent, cloud infrastructure and the capacity to set workflow standards. Therefore, AI investment concentration can become protocol-setting power where model access, compute infrastructure, cloud ecosystems and deployment standards are concentrated: actors with model capital and compute access may be better positioned to influence the rules by which future firms, markets and public systems coordinate action.

### 4.5. Compute-energy coupling: data-centre electricity demand as an infrastructure result

The compute-energy result is one of the clearest projection-based supports for the article's framework, but it must be interpreted cautiously. Global data-centre electricity consumption was approximately 415 TWh in 2024. Under the IEA Base Case, it reaches about 945 TWh by 2030 and about 1,200 TWh by 2035. The increase from 2024 to 2030 is 530 TWh, which means that demand is projected to become 2.28 times the 2024 level. The implied CAGR is 14.70% for 2024-2030. By 2035, the projected increase reaches 785 TWh, or 2.89 times the 2024 level, with an implied 2024-2035 CAGR of 10.13%. This is not interpreted as evidence that AI alone causes all data-centre electricity growth. Rather, AI-related compute expansion is treated as a major contributor to wider data-centre electricity pressure alongside cloud computing, storage, enterprise workloads, streaming and other digital infrastructure.

The significance of this result is methodological and policy-relevant. In the old production-function language, technology could be treated as an abstract productivity shifter. In the agentic economy, technology becomes an energy-bound action infrastructure. AI agents require inference and training infrastructure; robots require power, charging, maintenance and real-time control; protocols require continuous digital infrastructure. As a result, compute capacity and energy availability must be treated as internal variables of the model rather than as external background conditions.

**Table 4.5. Compute-energy pressure indicators**

| Compute-energy metric | Value | Calculation | Result interpretation |
|---|---|---|---|
| Observed data-centre electricity use | 415 TWh | Reported 2024 value | Compute already consumes electricity at the scale of a major economic infrastructure. |
| IEA Base Case 2030 projection | 945 TWh | Reported projection | Demand more than doubles by 2030 under the Base Case. |
| Absolute increase, 2024-2030 | 530 TWh | 945 - 415 | Additional demand becomes a grid-planning and energy-security issue. |
| Growth multiplier, 2024-2030 | 2.28x | 945 / 415 | AI and compute cannot be modelled without energy availability. |
| CAGR, 2024-2030 | 14.70% | (945/415)^(1/6)-1 | Annualised growth is far above ordinary mature-infrastructure growth. |
| IEA Base Case 2035 projection | 1,200 TWh | Reported projection | The energy constraint persists beyond 2030. |
| CAGR, 2024-2035 | 10.13% | (1200/415)^(1/11)-1 | Longer-window growth remains structurally significant. |

Source: Author's calculations based on International Energy Agency, Energy and AI, 2025.

This result also supports the article's claim that a country or firm cannot become fully agentic merely by adopting software tools. If data-centre capacity, electricity supply, cooling, grid connection and cyber-physical reliability are weak, then the practical action capacity of AI and robotic systems is constrained. The empirical result therefore connects digital transformation directly to industrial policy, energy security and infrastructure planning.

### 4.6. Robotic embodiment: stock-flow maturity and regional concentration

Robotics results show that the agentic economy is not only digital, while also requiring conceptual caution. In 2024, global industrial robot installations reached 542,076 units and the operational stock reached 4,663,698 units. The stock-flow ratio is 8.60. This means that the installed operational robot stock is about 8.6 times the annual installation flow. Industrial robots are treated not as full autonomous agents, but as persistent cyber-physical action capacity. In practical terms, they are not only annual capital expenditure; they form a durable stock of embodied action capacity that continues to execute production, logistics, inspection and other industrial tasks over time.

Annual installations represented approximately 11.62% of the global operational stock in 2024. This calculation should not be read as a depreciation estimate because the operational stock changes with retirements and reporting conventions. It is nevertheless useful as a scale indicator: the annual inflow of industrial robots is large enough to refresh and expand a multi-million-unit global stock. This supports the article's concept of robotic embodiment, according to which economic action increasingly moves into physical systems that consume energy, occupy space, require maintenance and generate safety and liability questions.

Regional concentration is also strong. Asia accounted for 74% of new industrial robot installations in 2024, Europe for 16%, the Americas for 9% and the remaining/rounding category for about 1%. The regional HHI is 0.5814, or 58.14 on a 0-100 concentration scale. This is a high concentration signal and indicates that future industrial competitiveness is not distributed evenly across regions.

**Table 4.6. Robotic embodiment and regional concentration indicators**

| Robotics indicator | Value | Calculation | Interpretation |
|---|---|---|---|
| Annual robot installations | 542,076 | Reported 2024 value | Robotic deployment is a mass industrial phenomenon. |
| Operational robot stock | 4,663,698 | Reported 2024 value | Robots constitute a persistent stock of cyber-physical capital. |
| Stock-flow ratio | 8.60 | 4,663,698 / 542,076 | Robots represent accumulated embodied action capacity, not only current-year investment. |
| Annual installations as share of operational stock | 11.62% | 542,076 / 4,663,698 | The annual flow is large relative to the existing stock. |
| Asia share of new installations | 74% | Reported regional share | Asia dominates new robotic deployment. |
| Europe share of new installations | 16% | Reported regional share | Europe remains significant but far behind Asia in annual deployment share. |
| Americas share of new installations | 9% | Reported regional share | The Americas account for a smaller share of new deployments. |
| Regional robot-installation HHI | 0.5814 | 0.74^2 + 0.16^2 + 0.09^2 + 0.01^2 | Robotic embodiment is geographically concentrated. |

Source: Author's calculations based on International Federation of Robotics, World Robotics 2025.

The robotics result has a direct implication for development economics. Low labour cost alone becomes a weaker competitive strategy when production systems increasingly depend on robotic throughput, automated quality control, sensor-driven logistics and cyber-physical coordination. Small and medium-sized economies must therefore treat robotics not only as a manufacturing issue but as an industrial-capability and education-policy issue. Without robotic literacy, maintenance capacity, safety standards and energy reliability, firms may remain users of imported automation rather than designers of agentic production systems.

### 4.7. Labour-market reallocation: work is repartitioned, not simply eliminated

The labour-market result supports a cautious reallocation interpretation rather than a simple labour-disappearance narrative. The World Economic Forum projects 170 million new roles and 92 million displaced or replaced roles by 2030, producing a projected net increase of 78 million roles. These values are projections, not realised outcomes. The projected job-structure shift is 22% of jobs. The ratio of new roles to displaced roles is 1.85, meaning that the projected number of new roles is about 1.85 times the projected number of displaced roles. At the same time, displaced roles equal 54.12% of new roles, which shows why the transition cannot be treated as socially painless.

The net increase of 78 million roles equals 45.88% of the projected new roles. This is important because the gross flows are much larger than the net result. A net-positive labour-market projection may still involve significant disruption if the new roles require different skills, locations, credentials and task architectures. The result therefore supports the article's concept of labour repartition: the key issue is not whether humans remain in the economy, but which tasks remain human-performed, which become agent-assisted, which are robot-executed and which require new forms of audit and sovereign judgement.

**Table 4.7. Labour-market reallocation calculations**

| Labour-market metric | Value | Calculation | Interpretation |
|---|---|---|---|
| Projected new roles by 2030 | 170 million | Reported WEF projection | Large job creation is expected under the projection. |
| Projected displaced/replaced roles by 2030 | 92 million | Reported WEF projection | Significant labour-market disruption remains. |
| Projected net change | 78 million | 170 - 92 | The projected net result is positive, but gross disruption is substantial. |
| Job-structure shift | 22% | Reported WEF projection | A large share of jobs is expected to be structurally affected. |
| New-to-displaced ratio | 1.85 | 170 / 92 | Projected creation exceeds displacement, but does not eliminate transition risk. |
| Displacement relative to new roles | 54.12% | 92 / 170 | More than half as many roles are displaced as created. |
| Net gain relative to new roles | 45.88% | 78 / 170 | The net result hides large two-sided labour flows. |

Source: Author's calculations based on World Economic Forum, Future of Jobs Report 2025.

These labour results strengthen the theoretical distinction between execution labour, coordination labour, information labour, audit labour and sovereign labour. Routine execution and structured information tasks face greater exposure to automation and agentic assistance. At the same time, audit labour and sovereign labour become more important because automated systems require verification, exception handling, escalation rules, human override and institutional accountability. The labour-market result therefore links directly to the article's claim that the human does not disappear from the economic system; the human function moves toward final responsibility, judgement and governance.

### 4.8. Mapping empirical findings to the agentic action-capacity framework

This article formulates the agentic economy as a system of action capacity rather than only output production. The empirical indicators in this Results section can be mapped to the action-capacity framework as follows. AI investment and AI adoption correspond to model/software-agent capacity. Robot installations and operational stock correspond to robotic or cyber-physical capacity. Data-centre electricity demand corresponds to compute-energy coupling. Labour-market reallocation corresponds to the transformation of human roles and the need for human sovereignty capacity. Protocolisation, auditability and verifiable trust are theoretically central but are not fully measured by the global statistics used here; they require sector-level data such as audit-log completeness, settlement latency, automated compliance records, appeal rates and override events.

**Table 4.8. Empirical mapping to the agentic action-capacity framework**

| Action-capacity variable | Empirical proxy used in this article | Measured result | Interpretive conclusion |
|---|---|---|---|
| M_t: model/software-agent capacity | AI investment, organisational AI use, EU/OECD firm AI adoption. | Corporate AI investment USD 252.3 billion; organisational AI use 88%; EU/OECD official firm adoption about 20%. | The software-agent layer is becoming an organisational and capital-allocation variable. |
| R_t: robotic/cyber-physical capacity | Robot installations and operational stock. | 542,076 installations; 4,663,698 operational stock. | Machine action is embodied in global industrial infrastructure. |
| C_t and En_t: compute and energy availability | Data-centre electricity consumption and IEA Base Case projections. | 415 TWh in 2024; 945 TWh projected by 2030. | AI economics must be connected to grid capacity, energy planning and infrastructure constraints. |
| H_t: human judgement and responsibility | WEF labour-market reallocation projection. | 170 million new roles, 92 million displaced, net +78 million. | Human labour is repartitioned toward supervision, audit, judgement and transition capability. |
| P_t: protocol quality | Not directly measured in the global dataset. | Requires sectoral data. | Future empirical work should measure programmable compliance, API dependence, automated settlement and rule-execution failures. |
| T_t: auditable trust | Not directly measured in the global dataset. | Requires sectoral data. | Future empirical work should measure log completeness, explainability, evidence quality, appeal mechanisms and audit access. |
| Omega_t: uncertainty and institutional constraints | Forecast caveats and source-definition differences. | Scenario-dependent. | The results support transition pressure, not deterministic causality. |

Source: Author’s synthesis based on the article’s action-capacity model and the empirical results of this section.

This mapping prevents the results from becoming a list of disconnected statistics. The same evidence base points toward a consistent structural conclusion: action capacity is no longer produced only by human labour, physical capital and ordinary technology. It is produced by the joint availability of model capacity, robotic capacity, compute, electricity, protocols, auditability and human governance. The empirical results do not yet measure all of these variables equally, but they already show that the measurable components are moving in the direction predicted by the theoretical framework.

### 4.9. Derived empirical synthesis without hypothetical numbers

To avoid artificial or unverifiable figures, the synthesis below does not introduce new assumed parameters. It uses only indicators calculated directly from the reported values in the empirical matrix. For this reason, the synthesis is better interpreted as a diagnostic dashboard than as a final econometric index. Its purpose is to show the relative strength of the transition across the measurable domains of the article.

Table 4.9. Derived empirical indicators from real statistical inputs

| Domain | Derived indicator | Calculated value | Empirical implication |
|---|---|---|---|
| AI diffusion | EU AI adoption CAGR, 2021-2025 | 26.95% | AI adoption is accelerating in official enterprise statistics. |
| AI diffusion | OECD AI adoption CAGR, 2023-2025 | 52.38% | OECD adoption more than doubled in two years. |
| AI organisational embedding | McKinsey organisational AI use, 2025 | 88% | AI is widespread in at least one business function among surveyed organisations. |
| AI capital | Generative-AI investment scale relative to corporate AI investment | 13.44% | Generative AI is a major investment sub-domain. |
| AI capital concentration | U.S.-to-China reported private AI investment ratio | 11.73x | AI capital allocation is geoeconomically concentrated. |
| Compute-energy coupling | Data-centre demand CAGR, 2024-2030 | 14.70% | Compute growth becomes an energy-system pressure. |
| Robotic embodiment | Robot stock-flow ratio | 8.60 | Robots represent accumulated embodied action capacity. |
| Robotic concentration | Regional robot-installation HHI | 0.5814 | New industrial robot deployment is regionally concentrated. |
| Labour reallocation | New-to-displaced roles ratio | 1.85 | The projection is net positive but highly disruptive. |

Source: Author's calculations. Each value is calculated directly from reported source data; no hypothetical figures are used.

The synthesis produces four main empirical findings. First, AI adoption is no longer a marginal digital phenomenon; it has become visible in official enterprise and firm statistics. Second, AI and robotics show different but complementary forms of agentic capacity: AI expands informational and decision-support capacity, while robots expand physical execution capacity. Third, the compute-energy result shows that the agentic economy has a material boundary: electricity, grid capacity and cooling are now part of the economics of AI. Fourth, the labour-market result shows that the transition is not captured by a simple replacement narrative; it is better understood as a reallocation of tasks, skills and responsibilities.

### 4.10. Hypothesis and proposition assessment

The hypotheses and theoretical proposition are assessed using the derived indicators above. The assessment is deliberately conservative. A hypothesis is described as supported only where the reported data and author calculations directly correspond to the claim. Where the evidence is projection-based or conceptually indirect, the result is described as supported with caution rather than fully confirmed.

Table 4.10. Hypothesis and proposition assessment based on real statistics and transparent calculations

| Hypothesis / proposition | Assessment | Evidence | Boundary of interpretation |
|---|---|---|---|
| H1: Firm-level AI adoption is accelerating. | Supported. | EU adoption CAGR = 26.95%; OECD adoption CAGR = 52.38%; 2025 | AI use is not automatically full agentic transformation; delegation and audit data |

| | | organisational AI use = 88%. | are required for that stronger claim. |
|---|---|---|---|
| H2: AI-related compute expansion contributes to wider data-centre electricity pressure. | Supported under IEA Base Case. | Data-centre electricity demand grows from 415 TWh in 2024 to 945 TWh by 2030; CAGR = 14.70%. | 2030 and 2035 values are projections, not realised facts. |
| H3: Robotic embodiment is already a global industrial reality, but industrial robots are treated as cyber-physical action capacity rather than full autonomous agents. | Supported. | 2024 installations = 542,076; operational stock = 4,663,698; stock-flow ratio = 8.60. | Industrial robots do not represent all service robots or all cyber-physical systems. |
| H4: Available labour-market projections are more consistent with a reallocation narrative than with a simple disappearance narrative. | Supported with caution. | WEF projects 170 million new roles, 92 million displaced roles and net +78 million roles; this remains projection-based evidence and not a realised labour-market outcome. | Projection does not guarantee fair transition; skill mismatch and distributional risks remain. |
| P5: Classical economic categories are under-specified for distributed socio-technical action. | Supported as conceptual proposition from empirical convergence. | Measured indicators correspond to model capacity, robotic capacity, compute-energy coupling and labour repartition. | This is not a single econometric coefficient; it is a theoretical proposition supported by multi-domain evidence. |

Source: Author's assessment based on Tables 4.1-4.9.

### 4.11. Direct answer to the research questions

The Results section answers the article's research problem in a disciplined way. The empirical evidence is sufficient to show that the preconditions of the agentic economy are measurable across several domains. AI adoption is accelerating; AI investment is large and concentrated; robotics has become a persistent stock of cyber-physical capital; compute demand creates a direct energy constraint; and labour-market projections show task reallocation rather than simple labour disappearance.

The results do not show that all firms or all countries have already entered a complete agentic economy. That stronger claim would require firm-level or sector-level evidence on delegated decision rights, automated execution, protocolised settlement, audit-log completeness, model-error costs and human override capacity. The correct conclusion is therefore neither sceptical nor exaggerated: the agentic economy is not yet a universal institutional order, but its measurable transition pressure is already strong enough to justify a distinct analytical vocabulary.

**Table 4.11. Research-question answer matrix**

| Research question | Results-based answer | Evidence used | What remains for future empirical work |
|---|---|---|---|
| RQ1: Are the measurable preconditions of the agentic economy already visible? | Yes, across AI adoption, investment, robotics, energy and labour-market indicators. | EU/OECD AI adoption; Stanford HAI investment; IFR robots; IEA data-centre demand; WEF labour projection. | Country-level and sector-level datasets should test the same framework with higher granularity. |

| RQ2: Which dimension shows the strongest transition pressure? | Compute-energy coupling and AI adoption acceleration show the clearest measurable pressure. | Data-centre electricity demand CAGR = 14.70% to 2030; EU/OECD adoption growth rates are high. | Energy-grid constraints and firm-level AI workflow data should be linked empirically. |
|---|---|---|---|
| RQ3: Is the agentic economy only a digital phenomenon? | No. Robotics data show that machine action is embodied in physical industrial systems. | 4.66 million operational industrial robots; 542,076 installations in 2024; stock-flow ratio = 8.60. | Service robots, logistics robots and autonomous systems should be added where reliable data exist. |
| RQ4: Does the transition eliminate labour? | The evidence supports labour reallocation rather than simple elimination. | WEF projection: 170 million new roles, 92 million displaced, net +78 million; this is projection-based evidence, not a realised labour-market outcome. | Task-level data should distinguish execution, coordination, information, audit and sovereign labour. |
| RQ5: Does the evidence justify a new economic vocabulary? | Yes, as a theoretical inference from multi-domain empirical convergence. | Model capacity, robotic capacity, compute-energy pressure and labour repartition are all visible in data. | Protocolisation and auditable trust must be measured through sector-specific operational data. |

Source: Author's synthesis from the Results section.

### 4.12. Calculation audit and reproducibility note

The calculations used in the Results section are intentionally simple and reproducible. The purpose is not to hide the empirical logic in a complex model, but to demonstrate transition pressure through transparent transformations of source values. The following formulas were used: absolute change = X_t - X_0; relative change = (X_t - X_0) / X_0; CAGR = (X_t / X_0)^(1/n) - 1; growth multiplier = X_t / X_0; stock-flow ratio = operational stock / annual installations; HHI = sum(s_i^2), where s_i is a country or regional share expressed as a decimal.

No values were simulated, imputed or assumed. This is a strength for a conceptual-quantitative article because it reduces the risk of unverifiable numbers. The limitation is that the results are diagnostic rather than causal. They identify acceleration, concentration and infrastructure pressure, but they do not estimate the causal effect of AI agents or robots on macroeconomic growth.

**Table 4.12. Calculation audit**

| Calculated indicator | Formula | Input values | Output |
|---|---|---|---|
| EU AI adoption CAGR | (20.0 / 7.7)^(1/4) - 1 | 7.7%; 20.0%; n=4 | 26.95% |
| OECD AI adoption CAGR | (20.2 / 8.7)^(1/2) - 1 | 8.7%; 20.2%; n=2 | 52.38% |
| Generative-AI scale | 33.9 / 252.3 | USD 33.9bn; USD 252.3bn | 13.44% |
| U.S.-to-China AI investment ratio | 109.1 / 9.3 | USD 109.1bn; USD 9.3bn | 11.73x |
| Data-centre demand multiplier to 2030 | 945 / 415 | 945 TWh; 415 TWh | 2.28x |
| Data-centre demand CAGR to 2030 | (945 / 415)^(1/6) - 1 | 415 TWh; 945 TWh; n=6 | 14.70% |
| Robot stock-flow ratio | 4,663,698 / 542,076 | 4,663,698; 542,076 | 8.60 |
| Robot regional HHI | 0.74^2 + 0.16^2 + 0.09^2 + 0.01^2 | 74%; 16%; 9%; 1% | 0.5814 |

| New-to-displaced roles ratio | 170 / 92 | 170 million; 92 million | 1.85 |
|---|---|---|---|

Source: Author's calculations.

### 4.13. Limitations of the Results section

The Results section has four limitations that should be stated explicitly. First, the study uses global aggregate indicators and does not estimate firm-level causality. Second, adoption indicators are not perfectly comparable because they use different populations, survey designs and statistical definitions. Third, the IEA and WEF values for future years are projections, not realised data. Fourth, protocolisation and auditable trust are theoretically central but not directly measured in the global sources used here.

These limitations do not weaken the article if they are framed correctly. They define the empirical scope of the study. The section demonstrates measurable transition pressure, not completed global transformation. Future empirical work should therefore operationalise protocolisation, auditability, coordination friction and human override capacity at sector level. Finance, logistics, energy systems, healthcare administration, public procurement and education platforms are especially suitable for this next stage because they generate operational records on latency, settlement, exception handling, model error, appeal and audit completeness.

### 4.14. Overall result of the empirical section

The overall empirical result is that the agentic economy is not yet a completed institutional order, but its measurable preconditions are already visible. The available data show four prominent pressures, while belonging to different evidence classes. AI adoption is accelerating in official firm statistics and is broadly reported in organisational surveys. AI capital allocation is large and concentrated. Robotics has reached a multi-million-unit global operational stock, meaning that non-human physical execution is already embedded in industrial systems. Compute demand is projected to place significant pressure on electricity systems, confirming that AI cannot be analysed separately from energy infrastructure.

The labour result is more complex and should not be simplified. The projection of a positive net job change does not remove the social cost of displacement, mismatch and reskilling. The correct result is not that AI and robotics will eliminate labour, but that they repartition labour into execution, coordination, information, audit and sovereign functions. This interpretation is consistent with the article's theoretical claim that the human remains central as the bearer of goals, responsibility, exception interpretation and legitimate override.

The quantitative results therefore support a disciplined version of the article's thesis: the agentic economy should be defined as a measurable transition pressure rather than as a completed empirical fact. The evidence is sufficient to justify a new analytical vocabulary, but not sufficient to claim that all firms, sectors or countries have already become agentic. The next stage of the research should move from global descriptive indicators to sector-specific datasets, especially in finance, logistics, energy and education, where coordination friction, auditability, protocol failure, latency, model error and human override can be directly measured.

## 5. Interpretation and Analysis

The Results section demonstrates that the agentic economy should be interpreted neither as a completed global order nor as a purely speculative concept. The clearest measurable pressures in the available data are AI adoption acceleration and projected compute-energy demand, but these indicators belong to different evidence classes and should not be treated as directly comparable magnitudes. Other relevant transition pressures include large AI capital-allocation signals, concentrated model and compute investment, multi-million-unit robotic embodiment and substantial labour-market reallocation. These indicators do not prove a single causal mechanism by themselves. They do, however, provide convergent evidence that the basic unit of economic analysis is changing from the human-centred firm using technology to a coordinated socio-technical system in which humans, AI agents, robots, protocols, compute infrastructure and energy constraints jointly shape economic action.

This chapter interprets those results as a structured diagnostic of economic transformation. The core analytical claim is that the economy is moving from a production-centred logic toward an action-capacity logic. In the traditional production function, technology is often compressed into a residual or an efficiency factor. In the agentic framework, technology becomes a partially acting infrastructure: it prepares decisions, executes tasks, coordinates workflows, verifies transactions, monitors risk and, in robotic form, acts physically in production and service environments. Therefore, the empirical findings should be read as evidence of redistribution of action rather than as evidence of simple automation.

The interpretation is deliberately conservative. The available global indicators measure adoption, investment, robotic stock, energy pressure and labour-market projections; they do not yet directly measure auditability, protocol quality, model-error cost, latency, override burden or the full composition of agentic workflows. For this reason, the chapter does not claim that the agentic economy has already replaced the old economy. Instead, it argues that the measured variables are strong enough to justify a new analytical vocabulary and a new empirical research programme.

### 5.1. Reading the Results as Transition Pressure, Not as Deterministic Proof

The most important interpretive step is to distinguish empirical transition pressure from deterministic transformation. The Results section contains three evidentiary classes: observed or reported data, institutional projections and author-calculated indicators. Observed or reported data include AI investment, enterprise AI use, robot installations and robot operational stock. Institutional projections include data-centre electricity demand under the IEA Base Case and labour-market reallocation by 2030. Author-calculated indicators include CAGR, relative growth, growth multipliers, stock-flow ratios and concentration measures. These derived values are transparent calculations, but they are not independent primary data.

This distinction raises the analytical quality of the article because it avoids the usual weakness in emerging-technology papers: confusing adoption signals with structural completion. A firm that uses AI in one function is not necessarily agentic. A country with AI tools is not necessarily prepared for autonomous or semi-autonomous economic coordination. A robot installation is not the same as full cyber-physical transformation of production. A labour-market forecast is not a guaranteed future. The evidence therefore supports a disciplined claim: measurable components of the agentic economy are already expanding, but the depth and institutional maturity of the transition must be assessed sector by sector.

The results answer the central research question in a qualified but strong way. The economy is not yet fully agentic, but the old categories of labour, capital, firm, market and technology are increasingly under-specified. They can still describe inputs, output and prices, but they cannot sufficiently describe who acts, through which model or protocol action is performed, how the action is verified, what infrastructure enables it, and who retains final responsibility. This is where the results become theoretically significant. Their contribution is not only statistical; it is categorical. They show why a new language is needed.

### 5.2. AI Adoption as Organisational Embedding: The First Layer of Agentic Capacity

The adoption results show that AI is no longer only a frontier research technology or a tool for isolated experimentation. Official statistics indicate that EU enterprise AI use increased from 7.7% in 2021 to 20.0% in 2025, while the OECD firm-level series increased from 8.7% in 2023 to 20.2% in 2025. The calculated CAGRs of 26.95% for the EU series and 52.38% for the OECD short-window series indicate rapid diffusion. At the same time, the organisational survey evidence reports a much higher level of use: 88% of surveyed organisations using AI in at least one business function in 2025. The gap between the official statistics and the survey statistic is not a contradiction; it is itself analytically meaningful. It shows that AI diffusion has several layers: formal enterprise adoption, firm-level use in official datasets, and broader organisational experimentation captured by surveys.

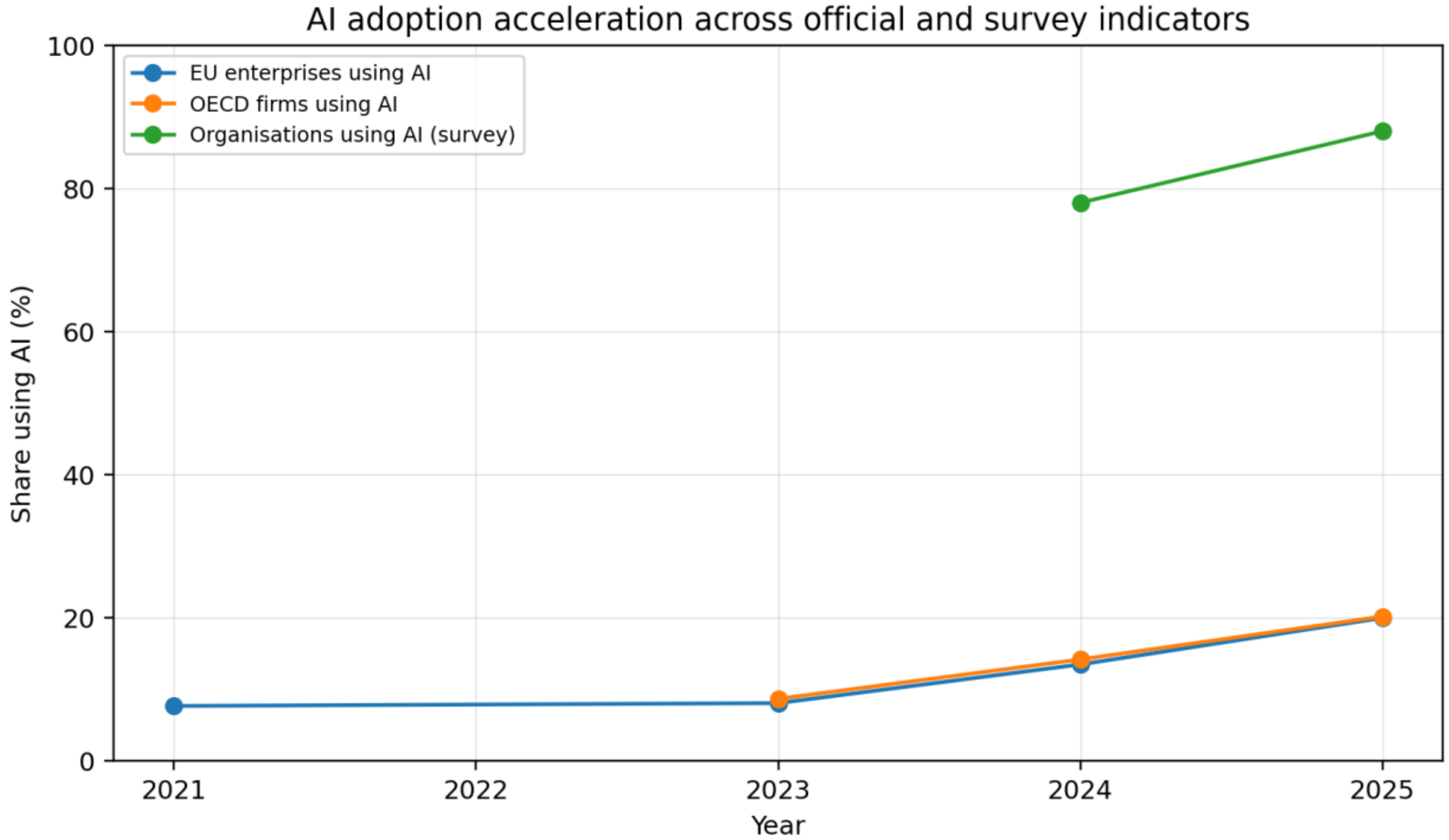

**Figure 5.1. AI adoption acceleration across official and survey indicators**
*Source: Author's visualisation based on the Results section values from Eurostat, OECD and McKinsey/Stanford HAI. Definitions differ by source and should not be merged into one single adoption index.*

Figure 5.1 shows that the diffusion of AI is not linear across measurement systems. The official EU and OECD series remain much lower than the organisational survey series because they measure different populations and definitions. This difference is important for interpretation. Official enterprise statistics are better for comparability and institutional measurement; survey evidence is better for detecting early workflow diffusion. Together, they show that AI is entering organisations before it is fully captured by conventional statistical systems.

The theoretical interpretation is that AI adoption is the first layer of agentic capacity. It does not automatically prove agentic transformation, but it creates the necessary precondition for it. Once AI enters marketing, sales, coding, administration, customer interaction, compliance, risk scoring or forecasting, the internal organisation of the firm begins to change. The firm is no longer only a hierarchy of human labour coordinated by managers. It becomes a mixed system where humans set goals, software systems process information, models generate recommendations, and organisational routines begin to depend on machine-readable workflows.

However, the distinction between AI use and agentic transformation must remain explicit. AI use means that a firm applies AI in at least one function. AI integration means that AI is embedded in recurring processes. Full agentic transformation begins only when agents, robots or executable protocols participate in preparation, execution, verification or coordination under auditable rules and human override. This distinction protects the article from overclaiming and makes the interpretation more acceptable for peer review. The result supports the emergence of agentic preconditions, not the completed arrival of agentic institutions.

### 5.3. AI Capital Concentration: From Technology Investment to Coordination Power

The investment results deepen the interpretation because they show that the agentic economy is not only an adoption phenomenon; it is also a capital-formation phenomenon. Corporate AI investment reached USD 252.3 billion in 2024, while private investment in generative AI reached USD 33.9 billion. Generative-AI investment therefore corresponds to 13.44% of the corporate AI investment figure used in the Results section. This should not be read as a strict accounting share because the categories are not identical. It is better understood as a scale comparison: generative AI has become a major investment sub-domain within the broader AI capital cycle.

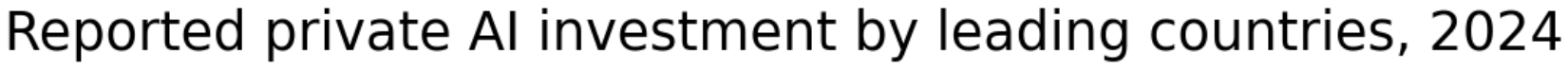

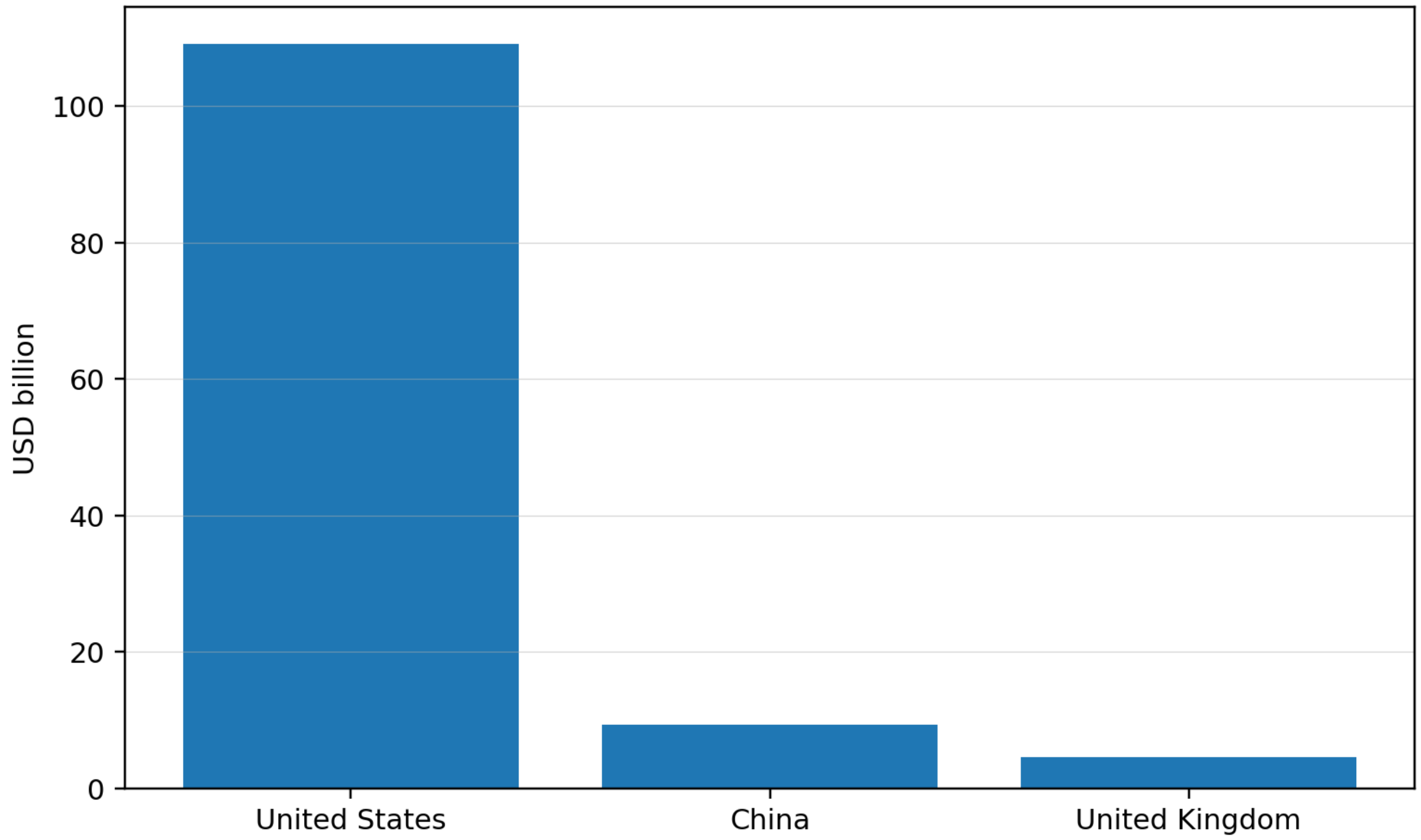

**Figure 5.2. Reported private AI investment by leading countries, 2024**
*Source: Author's visualisation based on Stanford HAI AI Index values used in the Results section.*

Figure 5.2 highlights a critical geoeconomic asymmetry. The reported U.S. private AI investment value of USD 109.1 billion is 11.73 times the China value and 24.24 times the UK value. Within the reported top-three country group, the United States accounts for 88.77%, and the top-three HHI is 0.7951. This HHI is not a global concentration index; it is a concentration measure for the reported top-three comparison. That boundary matters. Yet even with this limitation, the pattern is analytically strong: model capacity, compute access, talent concentration and cloud infrastructure are not evenly distributed.

The interpretation is that AI investment concentration may indicate emerging coordination-power asymmetries. In classical economics, investment expands productive capacity. In the agentic economy, AI investment may also expand the capacity to define how tasks are executed, how workflows are standardised, how APIs are integrated, how models are deployed, and how future organisational routines become dependent on external platforms or infrastructures. In this sense, AI capital can become institutional power where model access, compute infrastructure, cloud ecosystems and deployment standards are concentrated. The actors that control models, compute, data pipelines and deployment ecosystems may also shape the rules of economic coordination.

This result has a direct implication for smaller and developing economies. If they only import AI tools without building domestic capacity for data governance, model evaluation, compute access, technical education and audit institutions, they may become dependent users rather than strategic participants. The issue is not technological consumption; it is participation in rule-setting. The agentic economy rewards those who can design, supervise and audit systems, not merely those who can purchase subscriptions to them.

### 5.4. Compute-Energy Coupling: The Material Boundary of the Agentic Economy

The data-centre electricity result is one of the clearest empirical reasons why the agentic economy cannot be treated as a purely digital phenomenon. The Results section reports 415 TWh of data-centre electricity consumption in 2024 and an IEA Base Case projection of approximately 945 TWh by 2030 and 1,200 TWh by 2035. The calculated 2024-2030 growth multiplier is 2.28, with an implied CAGR of 14.70%. Over the longer 2024-2035 period, the multiplier reaches 2.89 and the implied CAGR is 10.13%. These figures support a central analytical conclusion: AI is materially constrained by electricity, grid capacity, cooling, location, energy prices and infrastructure reliability.

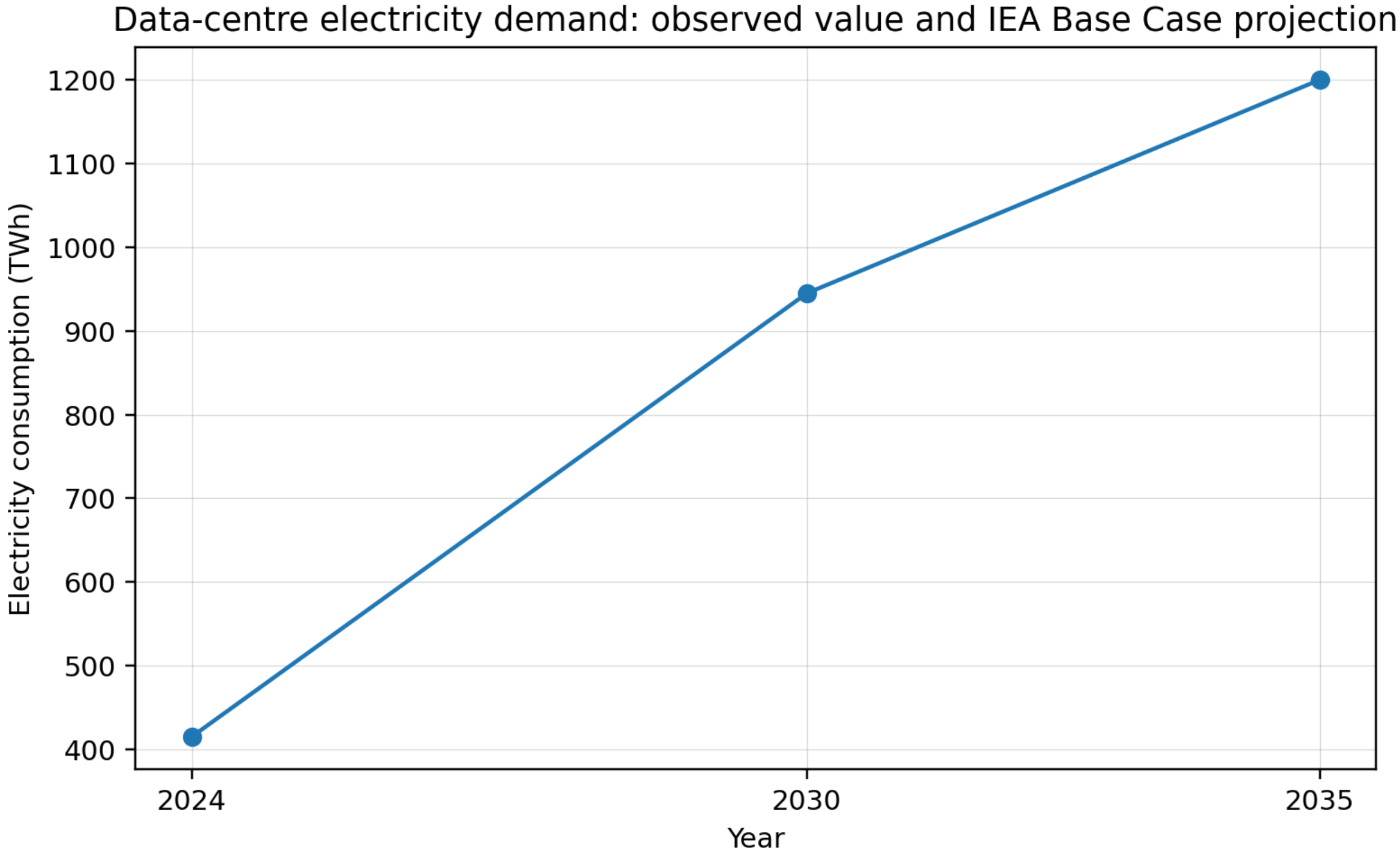

**Figure 5.3. Data-centre electricity demand: observed value and IEA Base Case projection**
*Source: Author's visualisation based on International Energy Agency values used in the Results section. Projection values are not realised facts and depend on IEA Base Case assumptions.*

Figure 5.3 changes the interpretation of digital transformation. If AI agents require continuous inference capacity, and if inference capacity requires electricity, then energy is not a secondary sectoral input. It becomes an internal variable in the economics of AI. The same applies to robotics: cyber-physical systems require power, charging, maintenance, sensor networks and real-time control. The agentic economy therefore depends on the coupling of compute and energy. A weak grid, unstable electricity supply, high connection delays or insufficient cooling infrastructure can constrain agentic capacity even where software adoption is high.

This finding also improves the theoretical model. In older production-function language, technology can be represented as a general productivity term. In the agentic action-capacity model, compute and energy must be separated. Compute represents the processing capacity through which models operate; energy represents the material condition that allows compute and robotics to function continuously. Treating them as separate variables prevents double counting and helps explain why two economies with similar AI adoption rates may differ sharply in real agentic capacity.

The policy interpretation is significant. Digital policy and energy policy can no longer be designed separately. Data-centre planning, grid investment, renewable integration, storage, demand response, cybersecurity and industrial strategy become part of the same economic architecture. A country that wants AI-enabled productivity without energy-system readiness may face bottlenecks that are invisible in conventional digital-readiness indicators. Therefore, the agentic economy is not only about algorithms; it is about infrastructure.

### 5.5. Robotic Embodiment: The Physical Extension of Economic Action

The robotics results show that the agentic economy extends beyond software. In 2024, global industrial robot installations reached 542,076 units and the operational stock reached 4,663,698 units. The calculated stock-flow ratio of 8.60 means that robots are not only current-year capital investment; they are a persistent accumulated stock of embodied action capacity. This is the physical counterpart of software-agent capacity. Software agents act in informational space; robots act in material space.

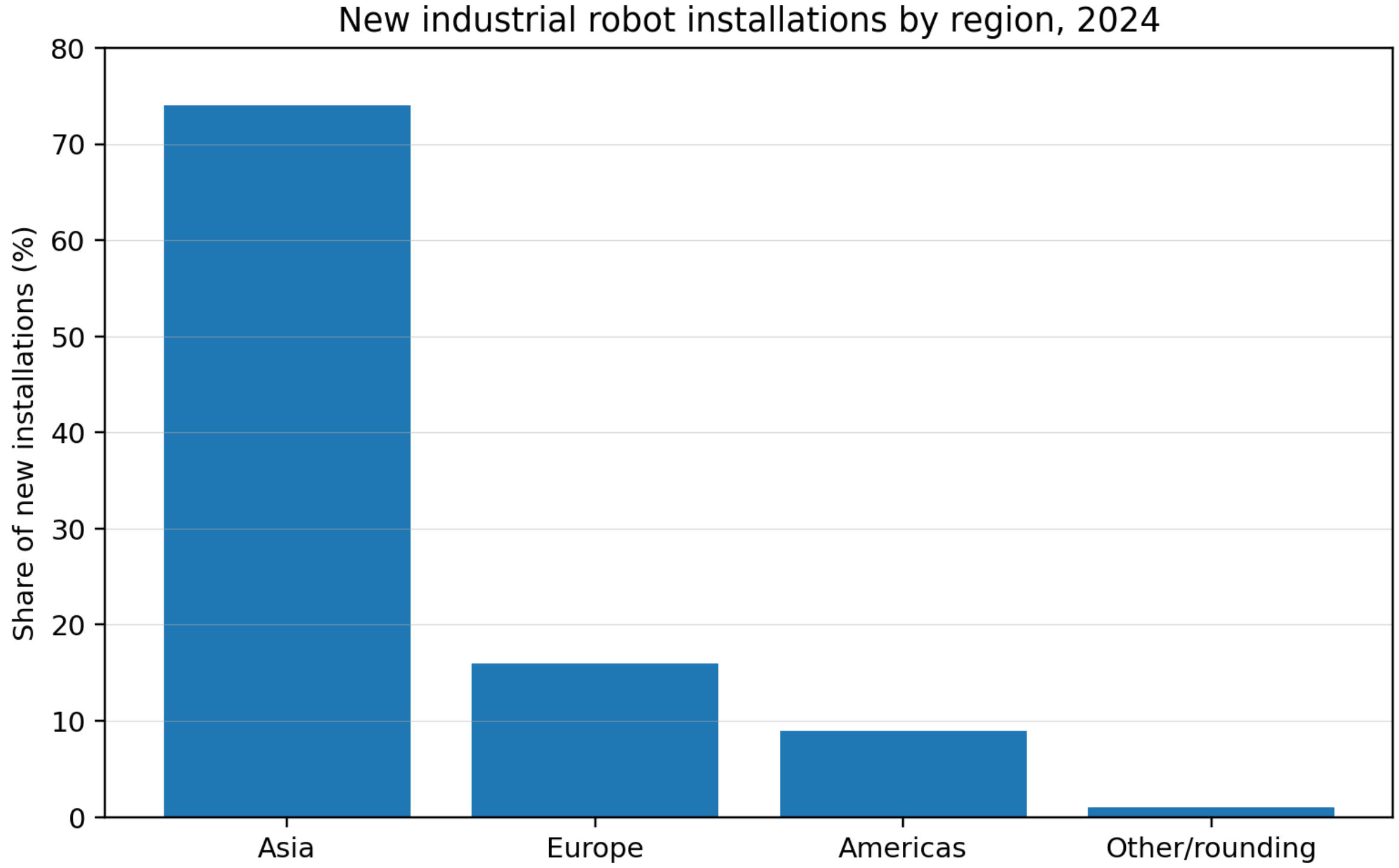

**Figure 5.4. New industrial robot installations by region, 2024**
*Source: Author's visualisation based on International Federation of Robotics values used in the Results section.*

Figure 5.4 shows the regional concentration of robotic embodiment. Asia accounts for 74% of new installations, Europe for 16%, the Americas for 9%, and the remaining/rounding category for approximately 1%. The calculated regional HHI of 0.5814 shows that robotic deployment is highly concentrated. This matters because industrial competitiveness is increasingly tied not only to labour cost or capital availability, but also to robotic throughput, maintenance capacity, safety standards, sensor infrastructure, energy reliability and cyber-physical coordination.

The interpretation is that robots transform the meaning of capital. A classical machine increases productivity, but it does not independently participate in action in the same way a connected robot does. A robot can move, sense, execute, stop, fail, require maintenance, interact with workers, create safety risks and generate liability. It therefore changes both the production process and the institutional environment surrounding production. Robotic capital is not merely physical capital; it is cyber-physical agency embedded in space.

This has strong implications for development strategy. The classical low-wage pathway becomes less reliable when global production systems can substitute routine physical execution with robotic throughput. Economies that lack robotics education, maintenance ecosystems, technical standards and industrial energy reliability may remain locked into lower-value segments of global production. The policy question is therefore not only whether firms can buy robots, but whether the whole institutional environment can support robotic action responsibly and productively.

### 5.6. Labour-Market Reallocation: Human Work Is Repartitioned, Not Abolished

The labour-market results require careful interpretation because they are often politically simplified. The WEF projection used in the Results section indicates 170 million new roles, 92 million displaced or replaced roles, and a net increase of 78 million roles by 2030, with a projected 22% job-structure shift. The new-to-displaced ratio is 1.85, while displaced roles equal 54.12% of new roles. The crucial point is that the net figure hides large gross flows. A positive net result does not mean a painless transition. It means that creation and displacement occur simultaneously and may affect different workers, regions, sectors and skill groups.

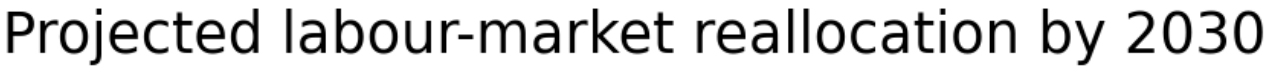

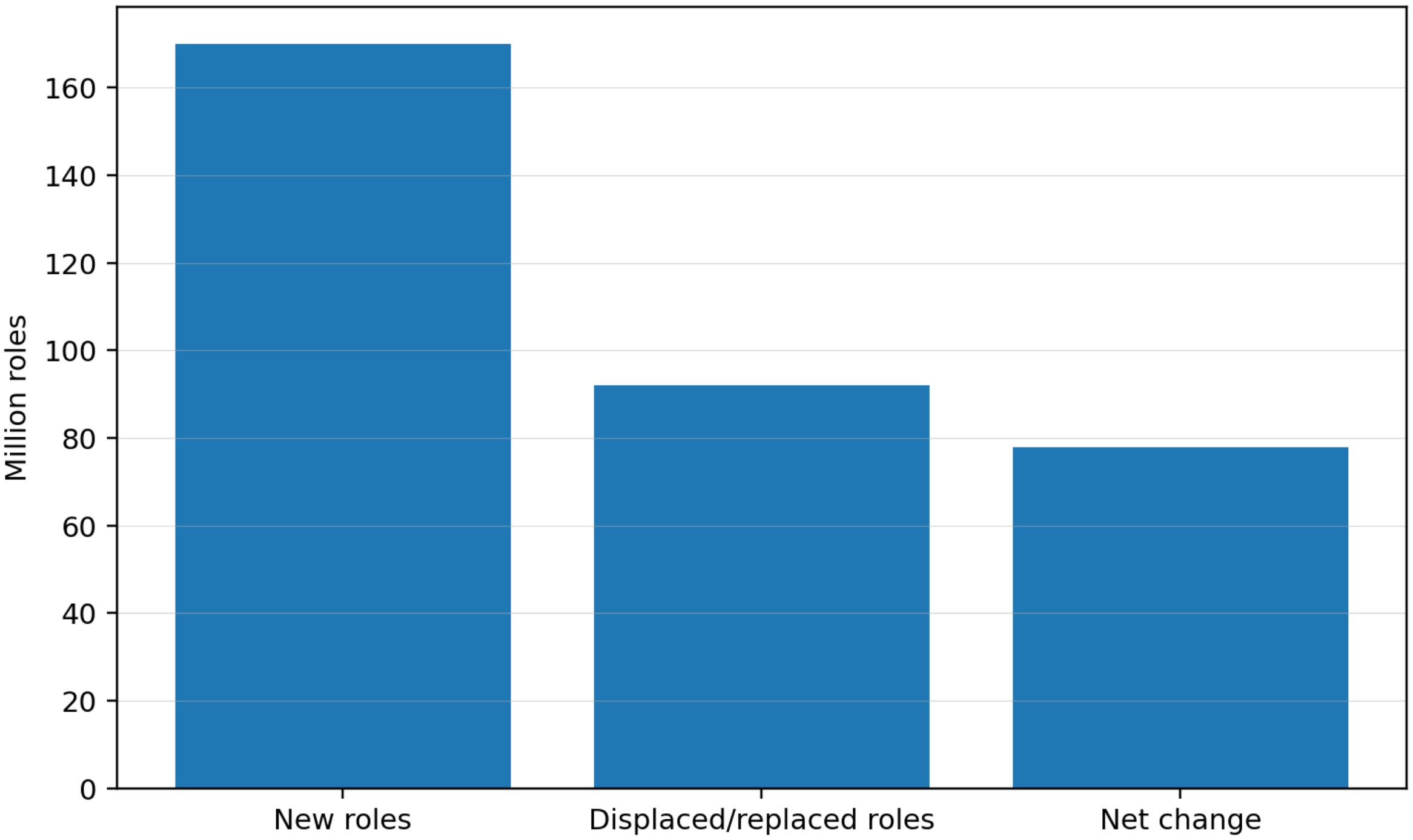

**Figure 5.5. Projected labour-market reallocation by 2030**
*Source: Author's visualisation based on World Economic Forum values used in the Results section. Values are projections, not realised outcomes.*

Figure 5.5 supports the article's central labour interpretation: the issue is not the disappearance of labour but the repartition of labour. Human work shifts from direct execution toward supervision, coordination, audit, exception interpretation and sovereign decision. Routine execution labour and structured information labour are more exposed to automation and AI augmentation. Audit labour, governance labour and sovereign labour become more important because automated systems must be checked, corrected, escalated and made accountable.

The concept of sovereign labour is therefore not rhetorical. It is the labour form that remains necessary when economic action is distributed across agents, robots and protocols. Humans define goals, determine normative boundaries, interpret exceptions, assess trade-offs, and exercise the right of override. If these functions are weakened or delegated without accountability, the agentic economy may become efficient but illegitimate. The labour result therefore links productivity, education and institutional legitimacy in one framework.

The policy implication is that reskilling must move beyond generic digital literacy. Workers need AI literacy, data reasoning, model-error detection, process orchestration, audit competence, cybersecurity awareness and domain-specific judgement. Universities and vocational institutions should not merely train people to use tools; they must train people to supervise systems that act. This is the difference between tool literacy and agentic literacy.

### 5.7. From Empirical Indicators to Action Capacity

The empirical findings become coherent when mapped onto the action-capacity framework. AI adoption and investment correspond to model/software-agent capacity. Robot installations and operational stock correspond to robotic or cyber-physical capacity. Data-centre electricity demand corresponds to compute-energy dependence. Labour reallocation corresponds to the transformation of human judgement and responsibility. Together, these results show that economic capacity is no longer generated only by labour, capital and general technology; it is generated by the coordinated availability of human authority, model capacity, robotic embodiment, protocols, compute, energy and auditable trust.

The action-capacity model is therefore not merely a theoretical formula. It is a way to organise empirical evidence. The Results section does not directly measure every variable in the model, but it measures enough core dimensions to justify the model's relevance. The missing variables are especially important: protocol quality and auditable trust. These require sector-level evidence such as programmable compliance records, automated settlement data, API failure rates, audit-log completeness, appeal mechanisms, model-error costs and human override records.

This boundary is a strength rather than a weakness if it is stated clearly. A strong article does not pretend that global public statistics can measure every dimension of an emerging agentic transition. Instead, it shows which dimensions are already measurable and which require future datasets. The present results establish the macro-diagnostic layer. The next empirical stage should move toward sector-specific and firm-level measurement of coordination friction, model error, latency, protocol failure, audit gaps and override burdens.

### 5.8. Coordination Friction as the Hidden Mechanism Behind the Results

The Results section does not directly estimate coordination friction, but the findings strongly indicate why the concept is necessary. AI adoption promises to reduce search cost, documentation cost, response time, analytical delay and some monitoring costs. Robotics promises to reduce execution delay, standardise physical processes and increase throughput. Protocolisation can reduce settlement time and verification disputes. However, each of these improvements can also create new frictions: model error, hallucination, cyber risk, API failure, audit opacity, energy bottlenecks, safety incidents and override delays.

This is where the article's contribution goes beyond a standard automation study. A conventional productivity interpretation might ask whether AI and robots increase output. The agentic interpretation asks whether the system can act with lower total coordination friction while preserving verifiability, safety and human legitimacy. This is a more demanding criterion. A system that accelerates decisions but increases audit opacity may improve short-run efficiency while increasing systemic risk. A robotised process that raises throughput but creates maintenance fragility or energy dependence may shift rather than remove friction. An AI-driven credit process that reduces processing time but increases opaque exclusion may lower administrative cost while raising institutional risk.

Therefore, the correct analytical move is to measure value not only as output growth but also as reduction in coordination friction, increase in verifiable trust, improvement in energy-adjusted efficiency and control of systemic risk. The Results section supports this move indirectly. AI adoption, robotics, energy demand and labour reallocation all show that the locus of value creation is moving toward coordinated action. The future empirical task is to quantify whether coordination becomes faster, cheaper, safer and more auditable.

### 5.9. Governance Implications: Human Sovereignty, Auditability and Institutional Design

The governance meaning of the results is that the agentic economy cannot be treated as a purely private-sector efficiency project. When AI systems shape workflows, robots perform physical tasks, compute infrastructure consumes increasing energy and labour markets undergo structural reallocation, the state, regulators, universities and firms all acquire new responsibilities. Governance must move from reactive regulation toward architecture-level oversight: standards for auditability, human override, model documentation, safety, data access, explainability, incident reporting and energy-system coordination.

Human sovereignty is the central normative condition. It means that humans must retain the capacity to define goals, supervise automated action, interpret exceptions and stop or correct systems when necessary. Human-in-the-loop design is not enough if the human is only a symbolic approver without time, competence or authority. Effective sovereignty requires institutional capacity: trained personnel, escalation protocols, audit trails, legal responsibility and governance structures that make override meaningful.

Auditability is equally important. In a human-centred economy, accountability could often be reconstructed through contracts, managerial responsibility and legal evidence. In an agentic economy, accountability also requires logs, model documentation, traceable decisions, evidence of data inputs, version control, protocol records and appeal mechanisms. Without these elements, the economy may become faster but less accountable. The Results section therefore implies that the expansion of AI and robotics must be matched by expansion of audit institutions.

For firms, the main governance implication is that adoption alone is not strategy. A firm may use AI widely but still lack agentic maturity if it cannot audit outputs, manage model drift, integrate human oversight, secure data pipelines or measure the costs of errors. For states, the implication is that competitiveness depends on compute access, energy reliability, robotics capacity, education and governance. For universities, the implication is that curricula must prepare graduates for agent management, audit, data reasoning and responsible oversight. The agentic economy is therefore an institutional project, not only a technological project.

### 5.10. Overall Analytical Conclusion of Chapter 5

The Results section provides a strong empirical-diagnostic basis for the article's thesis. AI adoption is accelerating in official and survey evidence; AI capital allocation is large and geographically concentrated; robotics forms a persistent stock of cyber-physical action capacity; data-centre electricity demand shows that compute is materially constrained by energy; and labour-market projections indicate structural reallocation rather than simple disappearance of work. These findings collectively support the interpretation that economic action is being redistributed across humans, AI agents, robots, protocols and infrastructure.

The strongest contribution of the analysis is not the claim that the agentic economy has already fully arrived. The stronger and more defensible claim is that the old economic vocabulary is now under-specified for the phenomena being measured. Labour must be expanded into execution, coordination, audit and sovereign labour. Capital must be expanded into model capital, compute infrastructure and robotic capital. Markets must be analysed not only as price systems but also as protocol-mediated coordination systems. Productivity must be supplemented by coordination-adjusted and energy-adjusted performance. Trust must be reconstructed as auditable and verifiable trust.

The interpretation also defines the article's empirical boundary. The data support transition pressure and conceptual extension, not a universal causal law. The chapter does not claim that every country, sector or firm has become agentic. It claims that the measurable preconditions of agentic transformation are now sufficiently visible to require systematic economic analysis. The next step should be to build sector-level datasets that measure coordination friction, auditability, protocolisation, model error, latency, energy bottlenecks and human override capacity.

In this sense, Chapter 5 completes the logical bridge between empirical results and theoretical contribution. The Results section shows what is changing; the interpretation explains why it matters. The agentic economy is not simply a new name for digitalisation, automation or platformisation. It is a broader coordination architecture in which action itself becomes distributed, infrastructure-bound, protocol-mediated and legitimacy-dependent. The central research implication is clear: future economic analysis must study not only how resources are allocated, but how coordinated action is produced, verified, governed and kept under human responsibility.

## 6. Conclusion

### 6.1. Synthesis of the article's argument

This article has developed the agentic economy as a conceptual and empirical framework for analysing a new stage of economic transformation. Its central claim is that the economy is not changing only because digital tools are spreading or because automation is increasing. It is changing because economic action itself is being redistributed across humans, AI agents, robots, protocols, compute infrastructures and energy systems. This redistribution affects how decisions are prepared, how tasks are executed, how transactions are verified, how systems are governed and where responsibility remains located.

The article does not argue that classical economic categories have become obsolete. Labour, capital, firm, market, productivity and trust remain necessary. The contribution is to show that these categories require extension. Labour must be understood not only as direct execution but also as coordination, audit, judgement and sovereign oversight. Capital must include model capital, compute infrastructure and robotic capital. Markets must be analysed not only as price systems but also as protocol-mediated coordination environments. Productivity must be supplemented by coordination-adjusted and energy-adjusted performance. Trust must be understood as auditable and verifiable trust.

The article's theoretical synthesis rests on two core concepts. The first is action capacity: the ability of a socio-technical system to select, execute, verify, correct and govern economic action. The second is coordination friction: the expanded field of costs that prevents economic action from being carried out efficiently and legitimately. These costs include traditional transaction costs, but also latency, model error, protocol failure, audit gaps, energy bottlenecks, cyber-physical risk and human-override burden. Together, these concepts provide a new language for analysing economies in which technology does not merely support action but increasingly participates in it.

### 6.2. Main empirical findings

The empirical findings support a disciplined and bounded conclusion: the agentic economy is not yet a completed global institutional order, but its measurable preconditions are already visible. The Results chapter shows that AI adoption is accelerating in official and organisational indicators; AI capital allocation is large and concentrated; industrial robotics has become a persistent stock of cyber-physical action capacity; data-centre electricity demand creates compute-energy pressure; and labour-market projections indicate structural reallocation rather than simple labour disappearance.

The first empirical finding concerns AI adoption. EU and OECD statistics show substantial growth in enterprise and firm-level AI use, while organisational survey evidence reports broader diffusion across business functions. These indicators do not prove full agentic transformation, but they show that AI is entering the internal structure of organisations. The relevant conclusion is that AI use is becoming a measurable organisational variable. This is the first layer of agentic capacity.

The second finding concerns AI investment concentration. The reported values for corporate AI investment and generative-AI investment indicate that AI is a broad capital-allocation signal. The leading-country comparison shows strong concentration, especially in the United States. The article interprets this not simply as investment asymmetry but as potential coordination-power asymmetry. Actors with stronger model capacity, compute access, data infrastructure and deployment ecosystems may influence the rules by which future economic processes are organised.

The third finding concerns compute-energy coupling. Data-centre electricity demand demonstrates that AI is not immaterial. Compute infrastructure depends on electricity, cooling, location, grid connection and energy security. This result is theoretically important because it moves energy inside the model of the digital economy. A firm or country cannot become fully agentic merely by adopting software tools if energy and compute infrastructure are insufficient.

The fourth finding concerns robotics. Industrial robot installations and operational stock show that machine action has a physical embodiment. Industrial robots are treated not as full autonomous agents, but as persistent cyber-physical action capacity; therefore, they are not only annual capital expenditure; they form a persistent operational stock that can execute tasks in material space. This supports the claim that the agentic economy is cyber-physical, not only digital. The regional concentration of robot deployment also indicates that future competitiveness will depend on robotic throughput, maintenance capacity, safety standards and industrial energy reliability.

The fifth finding concerns labour. The labour-market evidence does not support a simplistic claim that human work disappears. It supports a reallocation interpretation. Large projected flows of new and displaced roles indicate that the central question is how tasks, skills and responsibilities are redistributed. The article therefore proposes that labour should be analysed through execution labour, coordination labour, information labour, audit labour and sovereign labour. Human work remains central, but its function moves toward supervision, judgement, accountability and legitimate override.

### 6.3. Theoretical contribution

The article contributes to economic theory by integrating several literatures that are often treated separately: institutional economics, innovation economics, platform economics, labour economics, AI research, robotics, energy economics and AI governance. The integration is organised around the concept of action capacity. This concept allows the article to move beyond the question of whether technology increases output and toward the question of how economic action is produced, verified, governed and kept legitimate under distributed agency.

The proposed action-capacity framework extends the classical production function. In traditional language, technology is often represented as an efficiency factor. In the agentic framework, technology is decomposed into model/software-agent capacity, robotic capacity, compute capacity, energy availability, protocol quality and auditable trust. This decomposition is necessary because each component has different empirical indicators, institutional risks and policy implications. AI investment is not the same as robot stock; compute is not the same as energy; protocol quality is not the same as trust; and adoption is not the same as delegated agency.

The concept of coordination friction extends transaction-cost economics. Traditional transaction costs include search, bargaining, contracting, monitoring and enforcement. The agentic economy adds new frictions: latency, model error, protocol failure, audit opacity, cyber-physical risk, energy interruption and human-override burden. This extension is important because new technologies can reduce some old costs while creating new ones. A system may act faster but become less auditable; it may reduce human processing time but increase model-risk exposure; it may improve throughput while creating energy dependence. Therefore, the economic evaluation of agentic systems must measure total coordination quality, not only output or speed.

The article also contributes the concept of human sovereignty. Human sovereignty is not nostalgia for manual work. It is the institutional requirement that humans retain the capacity to set goals, interpret exceptions, define normative boundaries, contest outputs and stop or correct automated action. This concept connects economics with AI governance. An agentic economy without meaningful human sovereignty may generate efficiency but lose legitimacy. The article therefore treats human sovereignty as part of economic architecture, not as an external ethical supplement.

### 6.4. Methodological contribution

The methodological contribution is the development of a quantitative diagnostic design for an emerging agentic transition. The article does not force the data into a premature causal model. Instead, it uses recognised institutional sources and transparent transformations to test whether transition pressure is visible across multiple domains. This approach is appropriate because global harmonised microdata on agentic workflows, audit logs, protocol failures, model-error costs and override events are not yet available.

The methodology is intentionally reproducible. Each calculated result is derived from source-reported values using explicit formulas: percentage-point change, relative growth, compound annual growth rate, growth multiplier, scale ratio, stock-flow ratio and Herfindahl-Hirschman Index. No simulated or hypothetical values are introduced into the Results chapter. This strengthens the article because the calculations can be independently checked and replicated by future researchers.

The methodology also uses evidence classification. Observed or reported facts are separated from institutional projections, author-calculated indicators and theoretical interpretations. This prevents overclaiming. For example, data-centre electricity demand in 2030 is treated as an IEA Base Case projection, not as a realised fact. WEF labour-market figures are treated as projections, not as deterministic outcomes. AI adoption indicators from McKinsey, Eurostat and OECD are treated as convergence evidence, not as a merged panel dataset. These controls are essential for maintaining methodological discipline.

The diagnostic design is not a substitute for future causal research. It is a bridge toward it. By identifying which components of the agentic economy are already measurable and which remain unmeasured, the article provides a research architecture for future sector-level studies. The next stage should operationalise coordination friction, protocolisation, auditability, model error, latency, energy bottlenecks and human override capacity using firm-level or sector-level operational data.

### 6.5. Policy and governance implications

The findings have significant policy implications. First, digital policy and energy policy can no longer be separated. If AI and compute are energy-bound, then data-centre planning, grid reliability, renewable integration, storage, cooling infrastructure, electricity pricing and cybersecurity become part of AI strategy. Countries that focus only on AI adoption while neglecting energy infrastructure may face hidden constraints on real agentic capacity.

Second, industrial policy must incorporate robotics and cyber-physical readiness. Robot installations and operational stock show that automated physical execution is already part of global production. Competitiveness will increasingly depend on robotic maintenance, sensor infrastructure, technical education, safety standards and reliable industrial energy. Low labour cost alone is unlikely to remain a sufficient development strategy if advanced production systems are organised around robotic throughput and programmable logistics.

Third, labour-market policy must move beyond generic reskilling. The agentic economy requires AI literacy, data reasoning, model-error detection, process orchestration, audit competence, cybersecurity awareness and domain-specific judgement. Workers must be prepared not only to use tools but to supervise systems that act. Education systems should therefore treat agent management, auditability and responsible oversight as core economic competencies.

Fourth, regulatory policy must focus on auditability and accountability. AI and robotic systems that affect credit, employment, logistics, public services, compliance or safety require traceability, documentation, appeal mechanisms and meaningful human override. Governance must move from after-the-fact reaction to architecture-level design. Standards for logs, model documentation, incident reporting, protocol accountability and override authority should become part of economic regulation.

Fifth, competition policy must expand its object. Market power in the agentic economy may arise not only from prices or market share, but from control over models, compute, data pipelines, cloud infrastructure, APIs and protocol standards. This means that competition authorities and policymakers must understand coordination power, not only traditional monopoly power.

### 6.6. Implications for firms, universities and smaller economies

For firms, the article implies that adoption alone is not strategy. A company may use AI widely and still lack agentic maturity if it cannot audit model outputs, manage errors, secure data pipelines, integrate human oversight, measure latency or define escalation procedures. Agentic maturity requires organisational design: clear allocation of task rights, decision rights, verification duties and override responsibilities. Firms should therefore measure not only how many tools they use, but how reliably, safely and accountably those tools act within the workflow.

For universities, the implication is that education must prepare students for an economy in which part of action moves to machines while responsibility remains human. This requires a curriculum that integrates economics, data literacy, AI governance, systems thinking, ethics, robotics awareness, energy infrastructure and institutional analysis. The graduate of the agentic economy should not only know how to operate software; they should understand how to evaluate automated decisions, detect errors, supervise agents, interpret exceptions and preserve human sovereignty.

For smaller and developing economies, the article's conclusions are especially important. These economies may be tempted to treat AI as an imported software layer. However, the evidence suggests that agentic capacity also depends on compute, energy, data quality, robotics, education, audit institutions and protocol participation. If smaller economies remain only users of foreign models and platforms, they may become dependent on external coordination architectures. Strategic participation requires domestic capacity to evaluate models, build data infrastructure, regulate automated systems, train specialists and ensure energy reliability.

This does not mean every country must build frontier foundation models or massive robot industries. It means that every country must understand where it stands in the agentic value chain. Some may specialise in model evaluation, domain-specific AI, logistics automation, energy-efficient data infrastructure, robotics maintenance, audit services or governance standards. The policy task is to identify realistic positions within the agentic economy rather than assume that tool adoption equals transformation.

### 6.7. Limitations and future research agenda

The article has clear limitations. First, it uses global aggregate and institutional indicators rather than firm-level microdata. This means that the findings diagnose transition pressure but do not estimate causal effects. Second, the AI adoption indicators are not perfectly comparable because they come from different sources, populations and definitions. Third, the IEA and WEF values for future years are projections, not realised facts. Fourth, protocolisation, auditability and human override capacity are central to the theoretical framework but are not directly measured in the current global dataset.

These limitations define the next research agenda. Future studies should build sector-level datasets that measure coordination friction directly. Suitable sectors include finance, logistics, public procurement, healthcare administration, education platforms, energy systems and industrial robotics. These sectors generate operational records on settlement time, approval delays, disputes, failed matches, model errors, compliance exceptions, system outages, audit logs and override events. Such data would allow researchers to test whether agentic systems reduce total coordination friction or merely shift friction into less visible forms.

Future research should also distinguish between AI use, AI integration and full agentic transformation. Firm-level surveys should ask not only whether firms use AI, but which tasks are delegated, whether outputs are audited, whether decisions are contestable, whether humans can override automated action, and how model errors are recorded. Without these variables, adoption statistics will remain incomplete proxies.

Another future research direction is the measurement of human sovereignty capacity. This could include indicators such as AI-governance roles, employee training coverage, audit competence, escalation protocols, board-level oversight, appeal mechanisms and documented override procedures. Such indicators would make it possible to assess whether human responsibility remains real or becomes merely symbolic in automated systems.

A final direction is the development of country-level agentic readiness diagnostics. These should include AI adoption, compute access, electricity reliability, robot density, data-governance quality, audit institutions, education readiness, cybersecurity capacity and regulatory quality. Such diagnostics would allow policymakers to identify bottlenecks and avoid confusing digital consumption with agentic capability.

### 6.8. Final conclusion

The agentic economy is best understood as an emerging coordination architecture in which action becomes distributed, infrastructure-bound, protocol-mediated and legitimacy-dependent. The evidence reviewed and calculated in this article does not prove that this architecture is complete. It proves something more precise and methodologically defensible: its measurable preconditions are already visible across several global domains. AI adoption is accelerating; AI capital allocation is large and concentrated; robots form a persistent stock of embodied machine action; compute growth is tied to energy demand; and labour is being repartitioned across new task structures.

The article's final conclusion is that economic science must expand from an allocation-centred language to an action-capacity language. Resource allocation remains essential, but it is no longer sufficient. Future economic analysis must also ask who acts, through which system, under which protocol, with what energy and compute constraints, how the action is verified, and who retains final responsibility. These questions define the research frontier of the agentic economy.
The practical significance is equally clear. If policymakers, firms and universities treat the agentic economy merely as another phase of digital adoption, they will underestimate the importance of auditability, energy infrastructure, robotics, human sovereignty and institutional design. If they treat it as a completed transformation, they will overstate the evidence. The balanced position is stronger: the agentic economy is an emerging, measurable and strategically decisive transition pressure. It requires a new theoretical vocabulary, a transparent quantitative methodology and a future empirical programme focused on coordination friction, protocol quality, auditable trust and human override.

### Statement on the Use of AI Tools

The authors declare that BTUAI, an AI-assisted research and analytical support environment developed at Business and Technology University, was used in a limited supporting capacity during the preparation of this manuscript, with assistance from generative AI tools including ChatGPT. Its use was restricted to language editing, clarity improvement, translation support, structural refinement, formatting assistance, and alignment with academic writing conventions. The conceptual framework, research design, selection and interpretation of data sources, methodological decisions, calculations, arguments, and conclusions were developed, verified, and approved by the authors. The authors reviewed and edited all AI-assisted outputs and take full responsibility for the accuracy, integrity, originality, and final content of the manuscript.

### References

All references below include either a DOI link or an official/stable URL. The list is formatted in APA style and is restricted to sources relevant to institutional economics, innovation economics, platform economics, automation and labour, AI systems, robotics, compute-energy infrastructure, AI governance and the empirical source families used by the article.